\documentclass[reprint,
 amsmath,amssymb,
 aps,
pre, onecolumn]{revtex4-1}
\usepackage[utf8]{inputenc}
\usepackage[T1]{fontenc}
\usepackage{graphicx}
\usepackage{dcolumn}
\usepackage{bm}
\usepackage{hyperref}
\usepackage{mathptmx, amsmath, amssymb, mathrsfs}
\usepackage{physics}
\usepackage{braket}
\usepackage{xcolor}
\usepackage{mathrsfs}
\usepackage{euscript}
\usepackage{comment}
\newcommand{\ii}{\text{i}}

\newcommand{\mc}[1]{\mathcal{#1}}
\newcommand{\p}{\partial}

\DeclareMathOperator{\Prob}{Prob}
\DeclareMathOperator{\argmax}{argmax}
\begin{document}

\preprint{APS/123-QED}

\title{Generalized arcsine laws for a sluggish random walker with subdiffusive growth}

\author{Giuseppe Del Vecchio Del Vecchio}
\author{Satya N. Majumdar}%
\affiliation{%
 Universit\'e Paris-Saclay, CNRS, LPTMS, 91405, Orsay, France
}%

\date{\today}

\begin{abstract}

We study a simple one dimensional sluggish random walk model with subdiffusive growth. In the continuum hydrodynamic limit, the 
model corresponds to a particle diffusing on a line with a space dependent diffusion constant $D(x)= |x|^{-\alpha}$ and a drift 
potential $U(x)=|x|^{-\alpha}$, where $\alpha\ge 0$ parametrizes the model. For $\alpha=0$ it reduces to the standard 
diffusion, while for $\alpha>0$ it leads to a slow subdiffusive dynamics with the distance scaling as $x\sim t^{\mu}$ at late 
times with $\mu= 1/(\alpha+2)\leq 1/2$. In this paper, we compute exactly, for all $\alpha\ge 0$, the full probability 
distributions of three observables for a sluggish walker of duration $T$ starting at the origin: (i) the occupation time $t_+$ denoting the time spent 
on the positive side of the origin, (ii) the last passage time $t_{\rm l}$ through the origin before $T$, and (iii) the time 
$t_M$ at which the walker is maximally displaced on the positive side of the origin. We show that while for $\alpha=0$ all 
three distributions are identical and exhibit the celebrated arcsine laws of L\'evy, they become different from each other for 
any $\alpha>0$ and have nontrivial shapes dependent on $\alpha$. This generalizes the L\'evy's three arcsine laws for normal 
diffusion $(\alpha=0$) to the subdiffusive sluggish walker model with a general $\alpha\ge 0$. Numerical simulations are in 
excellent agreement with our analytical predictions.

\end{abstract}

\maketitle


\section{\label{sec:introduction}Introduction}

Since its discovery, Brownian motion remains a source of outstanding surprises and it is 
certainly the simplest and perhaps one of the most important continuous time stochastic processes 
\cite{Feller1950}. Its applications are countless and span across disciplines.
In one dimension, the position of a Brownian particle (also known as the Wiener process) evolves in time via the stochastic differential 
equation
\begin{equation}
    \frac{\dd x(t)}{\dd t} = \eta(t)
\end{equation}
where $\eta(t)$ is Gaussian white noise with zero mean and a correlator $\braket{\eta(t)\eta(t')} = 2D\, \delta(t-t')$ with 
$D>0$ being the diffusion constant. 
As is well known, this simple dynamics implies that the 
position of the Brownian particle is diffusive, i.e., the typical displacement grows as $x(t) \sim \sqrt{2D\, t}$ and 
that the positions
of the particle at two different times exhibit long range temporal correlation, i.e., $\braket{x(t)x(t')}=2D\, \min(t,t')$.

\begin{figure}
    \centering
    \includegraphics[width=0.7\linewidth]{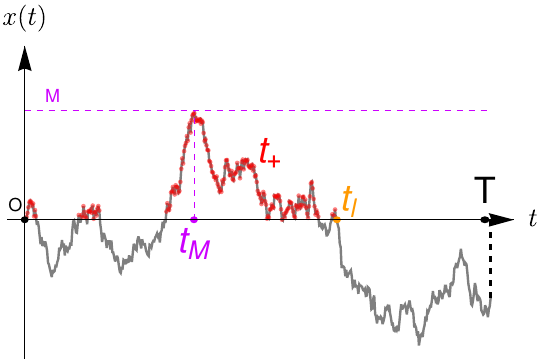}
    \caption{Sample path of a stochastic process 
observed in a window $[0,T]$ together with the three observables 
considered in this paper: $t_+$ in \eqref{eq:t_plus_def} is the sum of 
intervals on which the process is positive (red), 
$t_{\rm l}$ in \eqref{eq:last_passage_time_def} (orange) and 
$t_M$ in \eqref{eq:time_of_max_def} (purple). For completeness we also highlight the 
maximum value $M=x(t_M)$ reached by the process at time $t_M$ (purple). 
For Brownian motion the distributions of all three observables $t_+$, $t_{\rm l}$ and $t_M$ 
follow L\'evy's arcsine law \eqref{eq:arcsine_law}.}
    \label{fig:explanation_times}
\end{figure}

Consider a sample path of the Brownian motion of total duration $T$ starting at the origin $x(0)=0$ 
(see Fig. (\ref{fig:explanation_times})).
Three classical observables associated with such a path were introduced and studied by L\'evy~\cite{Levy1940}.
\begin{itemize}

\item {\bf Occupation time.} This is the total time spent by the Brownian motion of duration $T$ 
on the positive half axis
    \begin{equation}\label{eq:t_plus_def}
        t_+= \int_0^T \dd t \, \Theta(x(t)) 
    \end{equation}
    where $\Theta(x)$ is the Heaviside step function, i.e., 
$\Theta(x)=1$ for $x>0$ and $\Theta(x)=0$ for $x<0$.
Clearly $0\le t_{+}\le T$.
  
\item {\bf The last passage time to the origin.} This is the time instant $0\le t_{\rm l}\le T$ at which the Brownian path passes through the
origin for the last time before $T$, i.e.,  
    \begin{equation}\label{eq:last_passage_time_def}
        t_{\rm l} = \sup\{t\in[0,T]: x(t) = 0\}\quad .
    \end{equation}

\item {\bf The time of the maximum.}  This is the instant $0\le t_M\le T$ at which the particle achieves its
maximal displacement on the positive side, i.e., 
    \begin{equation}\label{eq:time_of_max_def}
         t_M = \underset{t \in [0,T]}{\argmax} \{x(t)\}\quad .
    \end{equation}

\end{itemize}
Fig. \ref{fig:explanation_times} shows a typical 
realisation of the Brownian path together with the three observables listed above.

These three observables are random variables each supported over
$[0,T]$. The marginal distributions of these random variables were
computed by L\'evy~\cite{Levy1940} (see also the book by
Feller~\cite{Feller1950}) and quite remarkably, they all share the
same probability distribution function (PDF) given by
\begin{equation}
p(t|T) = \frac{1}{T}f(t/T)\quad , \quad f(\xi) = \frac{1}{\pi\sqrt{\xi(1-\xi)}}\quad , \quad 0<\xi <1
\label{eq:arcsine_law}
\end{equation}
where $0\le t\le T$ generically stands for the value taken by each of the three 
observables $t_+$, $t_{\rm l}$ or $t_M$. Historically, the distribution \eqref{eq:arcsine_law} is often 
referred to as \emph{arcsine law} because the cumulative distribution function has the arcsine form, 
$\int_0^x f(\xi) \dd \xi = \frac{2}{\pi}\arcsin(\sqrt{x})$.

One striking feature of the PDF in Eq. (\ref{eq:arcsine_law}) is that it has a $U$ shape (convex)
with peaks at the two edges $t=0$ and $t=T$, and a minimum at $t=T/2$. Now the average value of any of the
three observables is $T/2$. The fact that the distribution is not peaked at its average value $T/2$ indicates
that their typical values (where the PDF is peaked) $t=0$ and $t=T$ do not coincide with the average $T/2$.
Thus, the probability density for a Brownian path to have values $t_+=0$ (where the path stays below the origin during
the whole interval $T$) or $t_+=T$ (where the path stays above the origin during the whole interval $T$) is
higher than its value at the average $t_+=T/2$. This indicates that the Brownian paths are `stiff', i.e., if they
make excursions to one side of the origin then, they tend to stick to that side typically.
In a fair game, for example, once one of the 
gamblers takes the lead, odds are that it keeps it for very long times \cite{dale1980}.

Despite the simplicity and the beauty of L\'evy's arcsine laws, their distributions $p(t|T)$ (for all
three observables) are not easy to derive by simple intuitive arguments. 
Different derivations exist in the literature that involve
either probabilistic arguments as in L\'evy's original derivation~\cite{Levy1940} or more
physics-oriented approaches using methods of  
path integrals (see, for example, the review~\cite{Majumdar2005}). Furthermore, 
because of the ubiquity of the Brownian motion, these three distributions have appeared in a wide range 
of seemingly very different situations and their applications lead to unexpected predictions 
about real world phenomena \cite{dey2022}. 
These features resulting from the original L\'evy's investigation of Brownian motion sparked a 
lot of interest in many different fields where first-passage properties
of a stochastic processes play a central role~\cite{Redner2001,Bray2013,Majumdar2010a},.
Consequently, all the three observables $t_+$, $t_{\rm l}$ and $t_M$ have been studied for a wide variety of
processes with diverse applications. 

For example, the distribution of the occupation time $t_+$
has been studied in a number of systems. These include random jump 
processes~\cite{Andersen1954,Watanabe1995},
renewal processes~\cite{Lamperti1958,Godreche2001,Burov2011} and 
a class of Gaussian Markov processes~\cite{Dhar1999,DeSmedt2001}.
It has also been studied in the context of persistence in many-body nonequilibrium 
systems~\cite{Dornic1998,Newman1998,Drouffe1998,Toroczkai1998,Baldassarri1999,Bray2013}, blinking nanocrystals~\cite{Brokmann2003,Margolin2005},
diffusion in a disordered environment~\cite{Majumdar2002a,Sabhapandit2006}, 
a class of spin glass models~\cite{MD2002}, anomalous subdiffusive processes~\cite{Bel2005,Barkai2006},
random acceleration process~\cite{RA2016}, 
fractional Brownian motion~\cite{Sadhu2018}, resetting Brownian motion~\cite{denHollander2019},
active run-and-tumble motion~\cite{SK2019,MS2023,MLS2024}, run-and-tumble  
process with stochastic 
resetting~\cite{Bressloff20} and
lattice random walk models in the presence of inhomogeneities~\cite{Radice2020,Kay2023}.

The last passage time $t_{\rm l}$ has also been studied in different contexts with multiple applications that include 
financial mathematics~\cite{dale1980,Nikeghbali2013}, models of phase noise in ring oscillators~\cite{Leung2004}, 
designing efficient 
numerical Monte Carlo methods \cite{Hwang2006, yu2021}, non-periodic inspection of stochastic systems in 
reliability engineering~\cite{Barkar2009}, 
random number generation \cite{robson2014}, models of 
competitive sports \cite{clauset2015} and noninteracting diffusive particles with connection to 
extreme value statistics \cite{Comtet2020}.

Finally the time $t_M$ at which a stochastic process achieves its maximum value in a given time interval
is an important observable studied extensively in the context of extreme value 
statistics~\cite{Majumdar2010a,Majumdar2020,Majumdar2024}.
It has an obvious application in finance: one would like to sell a stock at a time when the price of the stock (with
a certain maturity period $T$) is maximal and has been studied for a number of processes relevant in financial modelling
~\cite{Shepp1979,Buffet2003,Baz2004,Randon2007,MB2008,CB2014}. The distribution of $t_M$ was computed for a variety of
constrained Brownian motions (such as bridges, excursions, meanders) using an exact path decomposition method~\cite{MRKY2008}
and also by a real space renormalization group method~\cite{Schehr2010}. 
Another very curious and interesting exact relationship
has been found between the statistics of the perimeter and the area of the convex hull of a two dimensional
stochastic process and that of $t_M$ of one of the component processes in one dimension~\cite{RMC2009,MCR2010}. This mapping
has consequently been applied to study exactly
the statistics of convex hulls for several two dimensional processes including
Brownian motions~\cite{RMC2009,MCR2010,CBM15.1,CBM15.2}, random acceleration process~\cite{RMR2011},
branching Brownian motion with death~\cite{Dumonteil2013}, resetting Brownian motion~\cite{MMSS2021} and
run-and-tumble process in a plane~\cite{HMSS2020,SKMS2022}. The joint distribution of the time of the maximum
and that of the minimum has been computed exactly for a Brownian motion and Brownian bridge with interesting
applications to characterize the statistics of surface roughness in interface models in $(1+1)$-dimensions~\cite{MMS2019,MMS2020}.
The distribution of $t_M$ has also been computed for various other one dimensional processes such as
the random acceleration process~\cite{MRZ2010}, a class of anomalous diffusion processes~\cite{MRZ2010a},
vicious walkers problem~\cite{RS2011},
fractional Brownian motion~\cite{DW2016,Sadhu2018,AWW2020}, resetting Brownian motion in one dimension
~\cite{MMSS2021,SP2021} and heterogeneous diffusion process~\cite{Singh2020}. Finally, the distribution of $t_M$
was computed in the stationary state of a particle diffusing in a confining potential
and universal edge behaviors of the distribution of $t_M$ was uncovered~\cite{MMS2021,MMS2022}.
The symmety of the distribution of $t_M$ in a stationary time series was also proposed as an interesting
diagnostic to determine if the underlying dynamics
of the system is equilibrium or nonequilibrium~\cite{MMS2021,MMS2022}. 

From these extensive studies of the three observables $t_+$, $t_{\rm l}$ and $t_M$ for various processes mentioned
above, three clear facts have emerged: 
\begin{enumerate}

\item In general, for non-Brownian processes, there is no reason why
the PDF's of all three observables should be identical. 

\item The computation of the distribution
of each of these observables, whenever possible, require rather different techniques that are
generically nontrivial. 

\item There are very few non-Brownian processes for which the PDF's of
all the three observables can be computed exactly.

\end{enumerate}

The purpose of this paper is to present exact calculation of the PDF's of all the three observables
$t_+$, $t_{\rm l}$ and $t_M$ in a recently introduced {\em sluggish} random walk model~\cite{Zodage_2023}.
In this model, a single particle hops stochastically on the sites of a one dimensional lattice 
of lattice spacing $a$. The walker starts at the origin and
the dynamics takes place in discrete time with time step $\Delta t$.
In a single time step $\Delta t$, the particle
hops from site at $x$ to its right neighbour $x+a$ with probability 
$\left( \left(\frac{|x|}{a}\right)^{\alpha}+2\right)^{-1}$, to its left neighbour $x-a$ with
the same probability $\left( \left(\frac{|x|}{a}\right)^{\alpha}+2\right)^{-1}$,  
and stays at site $x$ with the remaining probability $1-2\, \left( \left(\frac{|x|}{a}\right)^{\alpha}+2\right)^{-1}$.
The exponent $\alpha\ge 0$ and the constant $2$ is just a short distance cut-off to ensure that the
hopping probabilities are well defined at the origin $x=0$ and the walk is completely symmetric around $x=0$. 
Even though the left/right hopping probabilities from a given site are identical, they are site-dependent and 
across a bond $(x, x+a)$
the particle has higher probability to move to the right (from $x$ to $x+a$) than to the left (from $x+a$ to $x$).
This asymmetry across a bond leads to an overall drift away from the origin.
This becomes more transparent when one writes down
the continuum space time equation by taking the limit $a\to 0$, $\Delta t\to 0$ but keeping
the ratio $a^{\alpha+2}/{\Delta t}$ fixed (this constant is chosen to be unity for convenience and amounts to a redefinition of units). In this
continuum limit, the Fokker-Planck equation
for the position distribution $P(x,t)$ evolves as~\cite{Zodage_2023}
\begin{equation}
\frac{\partial P(x,t)}{\partial t} = \frac{\partial}{\partial x}\left(D(x)\, \frac{\partial P(x,t)}{\partial x}\right)
+ \frac{\partial}{\partial x}\left(U'(x)\, P(x,t)\right)\, ,
\label{fp.1}
\end{equation}
where the space-dependent diffusion constant $D(x)$ and the drift potential $U(x)$ behave as
\begin{equation}
D(x)= \frac{1}{|x|^{\alpha}}\, , \quad {\rm and}\quad U(x)=\frac{1}{|x|^{\alpha}}\, .   
\label{du.1}
\end{equation}
Thus, the particle experiences a drift away from the center whose magnitude decays as a power law. Moreover,
the diffusion constant also decays as $|x|^{-\alpha}$ indicating that the dynamics becomes slower and slower (sluggish)
the further and further the particle moves away from the origin. This leads to a slow subdiffusive growth at late times
where the typical position scales as~\cite{Zodage_2023}
\begin{equation}
x\sim t^{\mu} \, \quad {\rm where}\quad \mu=\frac{1}{(\alpha+2)}\, .
\label{growth_law.1}
\end{equation}
For $\alpha=0$, this model reduces to the standard Brownian diffusion. For $\alpha>0$, this 
model represents a lattice random walk
in an inhomogeneous but non-random environment where at each site there is a `trap' of depth $l(x)$ that grows
logarithimcally with $|x|$ as $l(x)\sim \alpha \ln (|x|)$ for large $|x|$ and the hopping probability out of
the site at $x$ corresponds
to an Arrhenius escape probability from the trap~\cite{Zodage_2023}. 
It turns out that a similar hydrodynamic
equation for $P(x,t)$ (written down phenomenologically) was studied before~\cite{fa2003}, but its emergence
from a discrete lattice trap model was first pointed out in Ref.~\cite{Zodage_2023}. The position distribution
$P(x,t)$ can be solved exactly leading to a double peaked non-Gaussian shape with a singular `hole' (cusp) at 
$x=0$~\cite{fa2003,Zodage_2023} (see the next section for details). 
Let us also remark that a similar double peaked structure of the position
distribution with a hole at
the center was also found as the solution of the Fokker-Planck equation, 
$ \partial_t P(x,t)= \partial_x\left[ \sqrt{D(x)}\, \partial_x \left[ \sqrt{D(x)}\, P(x,t)\right]\right]$
in Refs.~\cite{cherstvy2013,SCT2023,SCT2023.1} with the same choice of the space dependent diffusion 
coefficient $D(x)=|x|^{-\alpha}$. This solution differs slightly in details from the solution
of the Fokker-Planck equation \eqref{fp.1} above~\cite{Zodage_2023}. Finally,
it was shown in Ref.~\cite{Zodage_2023} that
this sluggish walker model is solvable not just for the position distribution $P(x,t)$ , but for many other observables 
including the survival probability and
the distribution of the maximal displacement on the positive side--all having
nontrivial analytical forms.

In this paper, we show that the full probability
distribution of the three basic observables $t_+$, $t_{\rm l}$ and $t_M$ can
also be computed exactly for this sluggish random walk model, and each of them displays interesting
behaviors and deviations from the arcsine laws for all $\alpha>0$. Our analytical predictions are verified
by numerical simulations. Our result thus establishes the sluggish
random walk model as a rare, non-Brownian process for which the statistics of all the three
observables can be computed exactly.

The rest of the paper is organized as follows. In Section \ref{section:2} we present the sluggish random walk model
more precisely, recall the derivations of the free propagator and the survival probability and also
summarize the main results for the distributions of the three observables $t_+$, $t_{\rm l}$ and $t_M$. 
Section \ref{section:3} contains the bulk of the detailed technical calculations. We conclude in
Section \ref{section:4} with a summary and a brief outlook. Some details are relegated to the two
appendices.

\section{The model and a summary of the main results}
\label{section:2}

In this Section we first recall the precise definition of the sluggish random walk model introduced in 
\cite{Zodage_2023} and briefly outline the derivation of the propagator (or equivalently of the 
position distribution)
and the survival probability for general $\alpha\ge 0$, since these results
will be needed later for the derivation of the distributions of the three L\'evy observables $t_+$,
$t_{\rm l}$ and $t_M$. Since the actual derivations of these three distributions
are rather long and technical, we present a brief summary of the main results in this section.
The main technical part of the derivations are presented in detail in Section \ref{section:3}.

\subsection{Sluggish random walker}

Here we define the sluggish random walk model precisely and  
recall the computation of 1) the propagator in free space $P(x,t)$, i.e.,  the probability 
(density) that the walker is at $(x,t)$ given that it started at $x_0=0$ at time $t=0$ 
and 2) the survival probability $Q(x_0,t)$, i.e., the probability that
the walker, starting at $x_0>0$, does not cross the origin up to $t$.
As mentioned above, these two quantities will serve as basic ingredients in 
computing the probability distributions of $t_+$, $t_{\rm l}$ and $t_M$.

\begin{figure}
    \centering
    \includegraphics[width=0.8\linewidth]{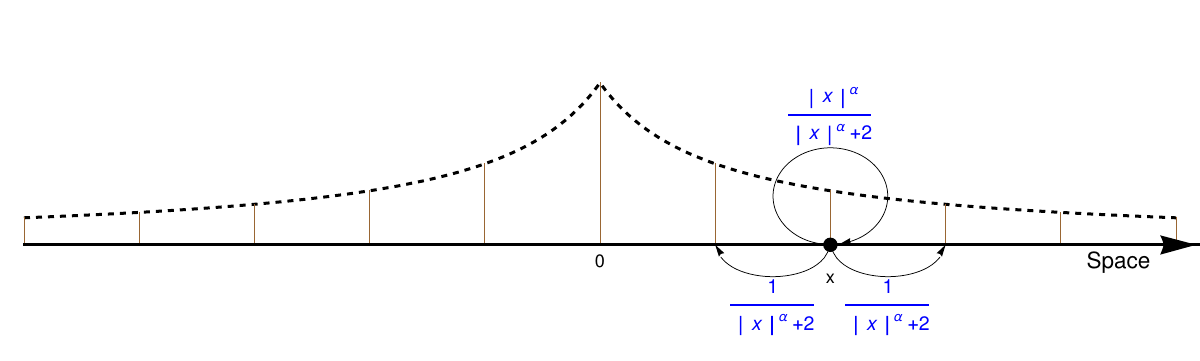}
    \caption{A sluggish random walker hopping on a one dimensional lattice (with lattice constant $a=1$) 
according to the dynamical rules \eqref{eq:sluggish_rules}. From a site at $x$, the walker moves to the left
neighbour $x-1$ with probability $1/(|x|^{\alpha}+2)$, to the right neighbour $x+1$ with probability
$1/(|x|^{\alpha}+2)$ and stays at site $x$ with the complementary probability $|x|^{\alpha}/(|x|^{\alpha}+2)$.
The dashed curve represents the repulsive potential $U(x)= 1/|x|^{\alpha}$ felt by the 
walker as given in \eqref{du.1}.}
    \label{fig:sluggish_random_walker}
\end{figure}

The model, as described briefly in the introduction, is very simple. A single particle
performs random walk on a lattice with lattice spacing $a$. The dynamics
occurs in discrete time steps each of duration $\Delta t$. In a single time step, the walker
at site $x$ either hops to one of its neighbours or stays at $x$ with the following
probabilities
\begin{equation}\label{eq:sluggish_rules}
    x(t+\Delta t)= \begin{cases}
        x(t)-a \quad&\text{with prob. ~} p_{x(t)\to x(t)-a}\\
         x(t)+a \quad&\text{with prob. ~}p_{x(t)\to x(t)+a}\\
          x(t) \quad&\text{with prob. ~} 1-\, \left( p_{x(t)\to x(t)-a}+ p_{x(t)\to x(t)+a}\right)\quad .
    \end{cases}
\end{equation}
where 
\begin{equation}
    p_{x\to x+a} = p_{x\to x-a}=\frac{1}{\left(\frac{|x|}{a}\right)^\alpha +2}, 
    \label{eq:Wx}
\end{equation}
with $\alpha \ge 0$. As mentioned earlier, the additional shift by $2$ in the denominator is just a short 
distance cut-off whose actual value is not important in the scaling limit.
A schematic representation of the sluggish dynamics of the walker is provided in Fig. 
\ref{fig:sluggish_random_walker}.

\subsubsection{Propagator}

The Master equation for the propagator $P(x,t|x_0,0)$, i.e., the probability that the 
particle is observed at $(x,t)$ when started from $x_0$ at $t=0$, can be written down easily using the 
dynamical rules \eqref{eq:sluggish_rules}
\begin{equation}\label{eq:master_equation}
    P(x,t+\Delta t|x_0) = \frac{1}{\left(\frac{|x-a|}{a}\right)^{\alpha}+2}\, P(x-a,t|x_0) + 
\frac{1}{\left(\frac{|x+a|}{a}\right)^{\alpha}+2}\, P(x+a,t|x_0) 
   + \frac{\left(\frac{|x|}{a}\right)^{\alpha}}{ \left(\frac{|x|}{a}\right)^{\alpha}+2}\, P(x,t|x_0)\, ,
\end{equation}
starting from the initial condition $P(x,t=0|x_0)=\delta_{x,x_0}$.
For notational convenience, we suppressed the initial time $t=0$ in $P(x,t|x_0)$. 
While one can solve this discrete space-time
recursion relation by generating function technique, to study the 
long distance late time behavior of the particle it is much
easier to study directly the continuum limit
where both the lattice spacing $a$ and the time step $\Delta t$ vanish
in an appropriate way. In this limit one can expand, for instance, $P(x+a,t|x_0)$ in a Taylor series in 
$a$ when $a\to 0$. Similarly, one can expand $P(x,t+\Delta t|x_0)$ in $\Delta t$.
Upon carrying out this expansion, it is easy to see that the appropriate continuum limit
is obtained when one takes the limit $a\to 0$, $\Delta t\to 0$ but with the ratio $a^{\alpha+2}/\Delta t$ 
kept fixed to a constant value. For the rest of this paper, we will choose this constant to be unity. 
The Master equation \eqref{eq:master_equation} then reduces to a forward 
Fokker-Planck equation in continuous space and time~\cite{Zodage_2023}
\begin{equation}
    \frac{\partial P(x,t)}{\partial t} = \frac{\partial^2}{\partial x^2} \left[\frac{1}{|x|^\alpha} P(x,t)\right]
    \label{eq:FPE}
\end{equation}
with the initial condition
\begin{equation}
    P(x,t=0|x_0) = \delta(x-x_0)\quad \, .
\end{equation}
This equation can be conveniently recast as \eqref{fp.1} to reveal the effective space-dependent diffusion
constant $D(x)$ and the drift potential $U(x)$ in \eqref{du.1}. Eq. (\ref{eq:FPE}) admits a scaling solution
at all times
\begin{equation}\label{eq:scaling_ansatz_propagator}
    P(x,t|x_0) = \frac{1}{t^\mu}\, H_P\left(\frac{x}{t^\mu}\right)
\end{equation} 
with 
\begin{equation}\label{eq:scaling_parameters}
    \mu = \frac{1}{\alpha+2}\, .
\end{equation}
The scaling function $H_P(z)$, symmetric in $z$, satisfies the second order differential equation for 
$z\ge 0$~\cite{Zodage_2023}
\begin{equation}
\label{HPz_eq}
    H_P''(z) + \left(\frac{z^{\alpha + 1}}{\alpha + 2} - 
\frac{2\alpha}{\alpha + 2}\frac{1}{z}\right)\, H_P'(z) + 
\left(\frac{\alpha(\alpha +1)}{z^2} + \frac{z^\alpha}{\alpha + 2}\right)\, H_P(z) = 0.
\end{equation}
The solution that is symmetric in $z$ and satisfies the boundary conditions $H_P(z) \to 0$ as $z \to \pm \infty$, is then 
given by~\cite{fa2003,Zodage_2023}
\begin{equation}\label{eq:propagator_free}
 H_P(z) = \frac{\mu^{1-2\mu}}{2\Gamma\left(1-\mu\right)}\, |z|^{\frac{1-2\mu}{\mu}}\, e^{- \mu^2|z|^{\frac{1}{\mu}}}\, ,
\end{equation} 
where we recall $\mu=1/(\alpha+2)$.
Thus the typical position of the walker scales subdiffusively as 
$x(t) \sim t^{1/(2+\alpha)}$ at late times for any $\alpha>0$.
For $\alpha=0$, one recovers the standard diffusion. The scaling function $H_P(z)$ in \eqref{eq:propagator_free}
has a non-Gaussian shape with two peaks located symmetrically away from the center $z=0$.
Moreover, it vanishes at the center as $\sim |z|^{\alpha}$ for all $\alpha>0$ indicating a singular 
`hole' at the center.
This origin of this `hole' can be traced back to the potential $U(x)= |x|^{-\alpha}$ that drives 
the particle away from the center.

\subsubsection{Survival probability}

We next consider the survival probability $Q(x_0,t)$ denoting the probability that starting at $x_0>0$,
the walker does not cross the origin up to $t$. It is easy to write down the backward Master equation
for $Q(x_0,t)$ treating the starting position $x_0$ as a variable. By examining where the particle
jumps in the first step, it is easy to see that for $x_0\ge 0$
\begin{equation}\label{eq:backward_discrete}
    Q(x_0,t+\Delta t) = \frac{1}{\left(\frac{|x_0|}{a}\right)^{\alpha}+2} \, Q(x_0-a,t) + 
\frac{1}{\left(\frac{|x_0|}{a}\right)^{\alpha}+2}\, Q(x_0+a,t) 
+ \left[1- \frac{2}{\left(\frac{|x_0|}{a}\right)^{\alpha}+2}\right] \, Q(x_0,t)\, ,
\end{equation}
with the absorbing boundary condition at the origin $Q(0,t)=0$ for all $t$ and the initial condition
$Q(x_0,0)=1$ for all $x_0>0$. In the continuum limit (where $a\to 0$, $\Delta t\to 0$
with $a^{\alpha+2}/{\Delta t}$ kept fixed to unity), it reduces to the backward Fokker-Planck
equation~\cite{Zodage_2023}  
\begin{equation}\label{eq:backward_survivial}
    \p_t Q(x_0,t) = \frac{1}{|x_0|^\alpha}\p_{x_0}^2 Q(x_0,t)\, .
\end{equation}
In Ref.~\cite{Zodage_2023} the solution was presented only for $\alpha=1$ for simplicity, 
but this can be trivially generalized
to arbitrary $\alpha\ge 0$. Eq. (\ref{eq:backward_survivial}) again admits a scaling solution of the
form
\begin{equation}\label{eq:scaling_ansatz_survival}
    Q(x_0,t) = H_Q\left(\frac{x_0}{t^\mu}\right)\, \quad {\rm with} \quad \mu= \frac{1}{\alpha+2}\, ,
\end{equation}
where the scaling function $H_Q(z)$ satisfies
\begin{equation}
    H_Q''(z) = - \frac{1}{\alpha + 2}|z|^{\alpha+1} H_Q'(z)\quad .
\end{equation}
The solution satisfying the required boundary conditions is 
\begin{equation}\label{eq:survival_free}
    H_Q(z) = 1-\frac{\Gamma \left(\frac{1}{\alpha+2},
\frac{|z|^{\alpha+2}}{(\alpha+2)^2}\right)}{\Gamma 
\left(\frac{1}{\alpha+2}\right)} = 1 - \frac{\Gamma\left(\mu, 
\mu^2|z|^{\frac{1}{\mu}}\right)}{\Gamma(\mu)}
\end{equation}
where $\Gamma(n,z) = \int_{z}^{\infty} \dd t \, e^{-t}t^{n-1}$ is the 
incomplete Gamma function. The scaling function has the asymptotic behaviors
\begin{eqnarray} \label{HQZ_asymp}
H_Q(z)\approx \begin{cases}
&  \mu^{2\mu - 1}|z| \hskip 2.5cm  {\rm as}\quad |z|\to 0 \\
& \\
&  1- \mu^{2(\mu-1)}|z|^{\frac{\mu-1}{\mu}}\,e^{-|z|^{\frac{1}{\mu}}} \quad {\rm as}\quad\quad |z|\to \infty\,, .
\end{cases}
\end{eqnarray}
The small $z$ behavior indicates from Eq. (\ref{eq:scaling_ansatz_survival}) that for $x_0\sim O(1)$, 
the survival probability decays at late times as a power law
\begin{equation}
Q(x_0,t) \sim \frac{|x_0|}{t^{\theta}} \, \quad {\rm with}\quad \theta=\mu= \frac{1}{\alpha+2}\, , 
\label{surv.1}
\end{equation}
where $\theta$ is known as the persistence exponent~\cite{Bray2013}.
Consequently, the first-passage probability to the origin also decays algebraically with time as 
\begin{equation}\label{eq:return_probability}
    F(x_0,t)= -\frac{dQ(x_0,t)}{dt} \sim t^{-\theta -1}\,  \quad \rm{with} \quad \theta=\mu\, .
\end{equation}

\subsection{Summary of the main results}

We now briefly summarize the main results for the distributions of the three observables $t_+$, $t_{\rm l}$
and $t_M$ for a sluggish random walker of duration $T$. The distributions for these three
observables for the lattice model in discrete time
are expected to approach a scaling limit when $T$ is large. We are interested only in this scaling limit
and the associated scaling functions can then be derived by analysing directly the
continuous hydrodynamic limit. 
Note that while in the discrete time lattice model,
the scaling forms of the distributions of $t_+$, $t_{\rm l}$ and $t_M$ are valid
only in the scaling limit when they are large, the total duration $T$ is large, with
the ratios $t_+/T$, $t_{\rm l}/T$ and $t_M/T$ fixed. However, for the continuous space time model,
these scaling forms hold for all times $T$. In this paper, we will analyse directly the model
in continuous space and time, and hence the scaling is expected to hold at all times.

\subsubsection{Distribution of the occupation time $t_+$} 

\begin{figure}
    \centering
   \includegraphics[width=0.7\linewidth]{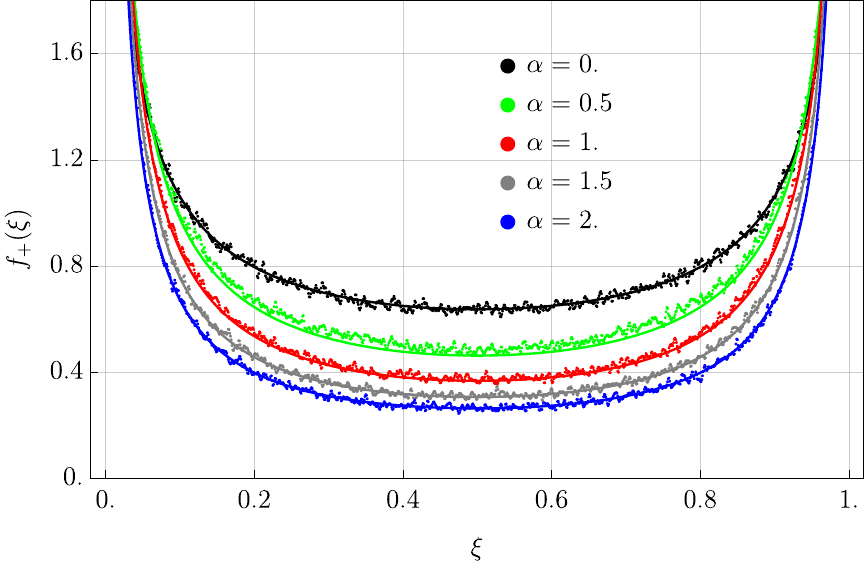}
    \caption{(Solid lines) Probability distribution $f_+(\xi)$ computed analytically in \eqref{eq:f_plus} for different
values of $\alpha$. 
(Points) Filtered numerical samples of the occupation time probability $t_+ = \int_0^T \, \dd \tau \, \Theta(x_\tau)$. 
We have simulated $10^6$ runs 
of the walker and took an observation time of $T=10^6$. }
    \label{fig:f_plus}
\end{figure}

As mentioned above,
for the lattice model evolving in discrete time, 
the PDF approaches the scaling form
\begin{equation}\label{eq:p_plus_scaling}
{\rm Prob.}\left[t_+=t|T\right]= p_+(t|T)\approx \frac{1}{T}f_+\left(\frac{t}{T}\right) \, ,
\end{equation}
in the limit when $t\to \infty$, $T\to \infty$ but with the ratio $t/T$ fixed.
However, for the continuous space time model that we analyze here, this scaling holds at
all times and the $\approx$ is replaced by $=$ in \eqref{eq:p_plus_scaling}.
We show that the exact scaling function is given by
\begin{equation}\label{eq:f_plus}
    f_+(\xi) = \frac{\sin(\pi \mu)}{\pi}\, \frac{\left[\xi\,(1-\xi)\right]^{\mu -1}}{\left[\xi^{2\mu} + 
2 \cos(\pi \mu)\, \left[\xi\, (1-\xi)\right]^{\mu}+ 
(1-\xi)^{2\mu}\right]}\, \quad {\rm with} \quad 0\le \xi\le 1\, \quad {\rm and}\quad \mu=\frac{1}{\alpha+2}\, .
\end{equation}
This distribution is $U$-shaped and symmetric around $\xi = 1/2$ implying that
$\braket{\xi}=1/2$ $\forall \alpha\in[0,+\infty)$. Since $\mu=1/(\alpha+2)<1$, the scaling function diverges at the two edges of the support:
$f(\xi)\sim \xi^{\mu-1}$ as $\xi\to 0$ and $f(\xi)\sim (1-\xi)^{\mu-1}$ as $\xi\to 1$.
For $\alpha=0$, i.e., $\mu=1/2$, it reduces to the standard arcsine form \eqref{eq:arcsine_law}
of the Brownian motion. 
We derive this result \eqref{eq:f_plus} by adapting
a backward Feynman-Kac approach previously used for Brownian motion~\cite{Majumdar2005} to the sluggish
random walk model. 
In Fig. \ref{fig:f_plus} we see a perfect agreement between our analytical formula and numerical simulations for different
values of $\alpha$. The data points matching the analytical prediction \eqref{eq:f_plus}
are obtained after application of a low-pass filter with
cutoff $0.001$ to the raw simulation of the microscopic dynamics rules by
\eqref{eq:sluggish_rules}. 

From the exact scaling function \eqref{eq:f_plus} one can compute all the moments of $\xi$. 
Instead of computing the integral $\int_0^1 \xi^m\, f_+(\xi)\, d\xi$ to compute the $m$-th
moment by using directly $f_+(\xi)$ from Eq. (\ref{eq:f_plus}), it turns out to be convenient
instead to first compute the Stieltjes transform of the scaling function $f_+(\xi)$ that
appears naturally within the backward Feynman-Kac approach. We obtain
\begin{equation}\label{eq:stieltjes_f_plus_result}
    \int_0^1 \dd \xi \frac{f_+(\xi)}{z-\xi}=R_+(z) + R_+(1-z)
\end{equation}
where
\begin{equation}
    R_+(z)=\frac{1}{z}\frac{1}{\left[1+\left(1-\frac{1}{z}\right)^{\mu}\right]}\quad . 
\end{equation}
Eq. (\ref{eq:stieltjes_f_plus_result}) holds for any $z$ in the complex plane excluding the cut
$[0,1]$ on the real axis.
Once we have the Stieltjes transform, the moments can be read off it by expanding around $z\to \infty$, i.e., by
using the identity
\begin{equation}
\int_0^1 \dd \xi \frac{f_+(\xi)}{z-\xi}= \sum_{m=0}^{\infty} \frac{1}{z^{m+1}}\, \int_0^1 \dd \xi\, f_+(\xi)\, \xi^{m}\, .  
\label{moment_exp.1}
\end{equation}
For example, the first few moments are given explicily by
\begin{equation}
\label{t+_moments}
        \braket{\xi^2} =  \frac{1}{2}-\frac{\mu}{4}\, , \quad
        \braket{\xi^3} =  \frac{1}{2}-\frac{3\mu}{8}\, , \quad
        \braket{\xi^4} = \frac{1}{2} - \frac{11 \mu}{24} + \frac{\mu^3}{48}\, , \quad    
        \braket{\xi^5} = \frac{1}{2} - \frac{25 \mu}{48} + \frac{5\mu^3}{96}\quad .
    \end{equation}

Let us also remark that this particular scaling function $f_+(\xi)$, parametrized by $\mu$, is actually well known as the
Lamperti distribution in the context of renewal processes~\cite{Lamperti1958}. Lamperti originally 
studied a renewal process where the time interval between successive renewals has a power law
tail $p(\tau)\sim \tau^{-1-\mu}$ with $0<\mu<1$. Next one assigns a signature $\pm 1$ to each interval
with equal probability $1/2$. Lamperti then
computed the distribution of the occupation time $t_+$, i.e., the total interval length during
which the process is positive
when the process runs for a total time $T$. He showed that if the average 
$\langle t_+\rangle=1/2$ (i.e., the positive and negative
signatures are equally likely), then the distribution $p_+(t|T)$ of $t_+$ has the
scaling form \eqref{eq:p_plus_scaling} with the scaling function precisely given by \eqref{eq:f_plus}.
Indeed, this Lamperti form has appeared in numerous other models essentially due to the fact
that many such processes can be recast as a renewal process with an effective persistence 
exponent $\mu$~\cite{Baldassarri1999,Bel2005,Radice2020}. Indeed, in our model, the time between two
successive returns to the origin has a power law tail $\sim \tau^{-1-\mu}$ for large $\tau$ (see 
Eq. (\ref{eq:return_probability})).
Since the successive returns to the origin are statistically independent and positive and negative excursions are
equally likely, we can indeed apply
Lamperti's theorem to the sluggish random walker problem to obtain 
the result \eqref{eq:f_plus}. This then provides an alternative confirmation of our exact result for $p_+(t|T)$.

\subsubsection{Last passage time}

\begin{figure}
    \centering
    \includegraphics[width=0.7\linewidth]{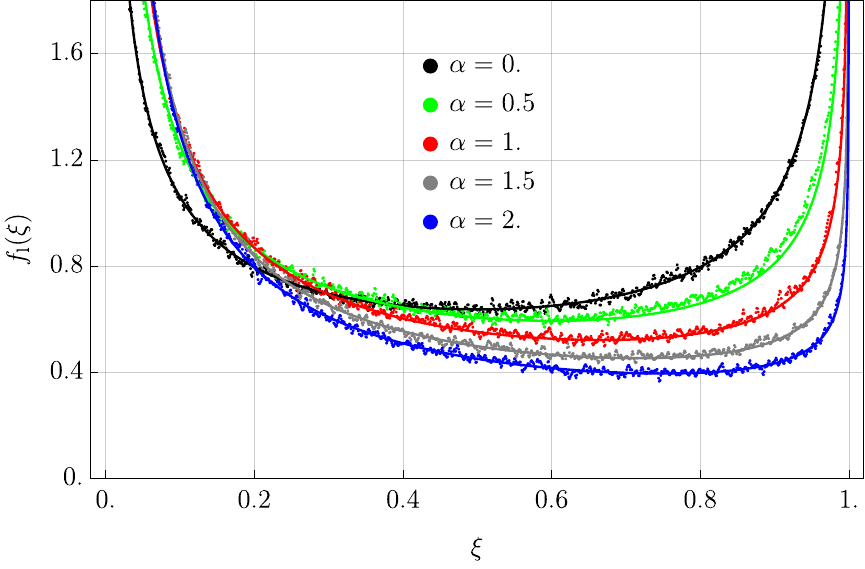}
    \caption{(Solid lines) Probability distribution $f_{\rm l}(\xi)$ computed analytically in \eqref{eq:f_l} for
different values of $\alpha$. 
(Points) Numerical obtained scaling function $f_{\rm l}(\xi)$. 
Simulation details are same as in Fig. \ref{fig:f_plus}. }
    \label{fig:f_l}
\end{figure}

For the last passage time $t_{\rm l}$ through the origin for the sluggish random walk model 
with parameter $\alpha\ge 0$,
we show that at late times
the PDF of $t_{\rm l}$ reaches the scaling form
\begin{equation}\label{eq:p_l}
{\rm Prob.}\left[t_{\rm l}=t|T\right] = p_{\rm l}(t|T) \approx \frac{1}{T}\, f_{\rm l}\left(\frac{t}{T}\right)\, ,
\end{equation}
valid in the scaling limit $t$ large, $T$ large but with their ratio $\xi=t/T$ fixed. The exact scaling function is given by
\begin{equation}\label{eq:f_l}
    f_{\rm l }(\xi) = \frac{\sin(\pi\mu)}{\pi}\frac{1}{\xi^{1-\mu}(1-\xi)^\mu}\quad  {\rm with} \quad 0\le \xi\le 1\, \quad {\rm and}\quad \mu=\frac{1}{\alpha+2}\, .
\end{equation}
As in the case of the occupation time, for the continuous space time sluggish model, the $\approx$ is replaced by $=$ in Eq. (\ref{eq:p_l}).
For $\alpha=0$, it again reduces to the arcsine form \eqref{eq:arcsine_law}. However for any $\alpha>0$,
unlike the occupation time distribution, the scaling function $f_{\rm l}(\xi)$ is 
evidently asymmetric around $\xi=1/2$. It diverges as both edges $\xi\to 0$ and $\xi\to 1$,
but with different exponents. As $\alpha$ increases, i.e., $\mu=1/(\alpha+2)$ decreases,
the divergence near $\xi\to 0$ becomes stronger than at the other edge $\xi\to 1$.
This is intuitively easy to understand, because once the walker crosses the origin at small
times, it strays away from the origin due to the repulsive potential $U(x)=|x|^{-\alpha}$ 
and becomes increasingly sluggish to recross the origin. Consequently the event
$\xi\to 0$ has more probability weight than the event $\xi\to 1$.
In Fig. (\ref{fig:f_l}) we compare our analytical prediction with numerical simulations
for different values of $\alpha$ finding excellent agreement.
The moments of the scaled last passage time $\xi=t_{\rm l}/T$ are easily found by direct integration of \eqref{eq:f_l} 
\begin{equation}
\label{tl_moments}
    \braket{\xi^m}= \frac{\sin(\pi\mu)}{\pi}\frac{\Gamma(1-\mu)\Gamma(\mu + m)}{\Gamma(m+1)\, .}
\end{equation}
The moments are increasing 
functions of $\mu$ (i.e., decreasing functions $\alpha$).  This behavior is exactly opposite to 
the moments of $t_+$ in \eqref{t+_moments} that decrease with increasing $\mu$. 

\subsubsection{Time to reach the maximum}

Turning now to the observable $t_M$ in \eqref{eq:time_of_max_def}, i.e., the time at which the sluggish walker
of duration $T$ achieves its maximum displacement on the positive side, we show that its PDF again takes
a scaling form in the limit 
\begin{equation}
\label{tM_scaling.1}
{\rm Prob.}\left[t_M=t|T\right] = p_M(t|T) \approx \frac{1}{T}\, f_M\left(\frac{t}{T}\right)\quad \, ,
\end{equation}
in the limit when both $t$ and $T$ are large with their ratio $t/T$ fixed.
Once again, the $\approx$ is replaced by $=$ in Eq. (\ref{tM_scaling.1}) for the continuous space time model.
The scaling function $f_M(\xi)$ is however much less explicit compared to the two previous cases namely $f_+(\xi)$
and $f_{\rm l}(\xi)$. It turns out that a more natural object is the Stieltjes transform of $f_M(\xi)$ rather
than $f_M(\xi)$ itself. In fact, the Stieltjes transform emerges quite naturally in our method and we obtain
\begin{equation}\label{eq:stieltjes_time_of_max}
    \int_0^1 \dd \xi\, \frac{f_M(\xi)}{z-\xi} = \frac{\mc{N}}{z}\int_0^{\infty} 
\frac{\dd x}{\left[g_\mu\left(\sqrt{1-\frac{1}{z}}\, x\right)\right]}\, 
\frac{g_\mu'(x)}{g_\mu(x)}\, ,
\end{equation}
where 
\begin{equation}\label{eq:normalisation_M}
    \mc{N} = \frac{2^{\mu}}{\Gamma(1-\mu)}
\end{equation}
is a normalisation constant ensuring $\int_0^1 \dd \xi \,f_M(\xi) = 1$ and the function $g_\mu(x)$
is given by 
\begin{equation}\label{eq:auxiliary_function_g}
    g_\mu(x) = x^\mu\left[I_\mu(x) + I_{-\mu}(x)\right]
\end{equation}
with $I_\mu(x)$ denoting the standard modified Bessel function of index $\mu$~\cite{watson_treatise_bessel}.
For the Brownian case where $\alpha=0$ and hence $\mu=1/2$, this function has a simple expression
\begin{equation}
g_{1/2}(x)= \sqrt{\frac{2}{\pi}}\,\, e^x\, .
\label{gmux_Brownian}
\end{equation}
For general $0<\mu<1/2$, the function $g_\mu(x)$ has the asymptotic behaviors
\begin{eqnarray}
\label{gx_asymp}
g_\mu(x) = \begin{cases}
& \frac{2^{\mu}}{\Gamma(1-\mu)}+ \frac{2^{-\mu}}{\Gamma(1+\mu)}\, x^{2\mu}+\ldots\,  \quad\quad\quad\quad\,\,\, {\rm as} \quad x\to 0 \\
& \\
& \sqrt{\frac{2}{\pi}}\, x^{\mu-1/2}\, e^{x}+\ldots\, \quad\quad\quad\quad\quad\quad\quad {\rm as} \quad x\to \infty\, .
\end{cases}
\end{eqnarray}
 
From the explicit Stieltjes transform in \eqref{eq:stieltjes_time_of_max}, one can 
compute the moments by expanding around $z\to \infty$ as in
Eq. (\ref{moment_exp.1}). For example, the first two moments are given by
\begin{eqnarray}
\label{fM_moments}
    \braket{\xi} &= & \frac{\mc{N}}{2}\, \int_0^{\infty}\dd x\,\frac{x\, [g'_\mu(x)]^2}{[g_\mu(x)]^3} \\
    \braket{\xi^2} &= & \frac{\mc{N}}{8}\,\int_0^{\infty} \dd x\,\frac{g_\mu'(x)}{[g_\mu(x)]^2}\, \Bigg[-x^2 g_\mu(x) g_\mu''(x)
+2\, x^2\, [g_\mu'(x)]^2+x\, g_\mu(x)\,g_\mu'(x)\Bigg]\, ,
\end{eqnarray}
where $g_\mu(x)$ is given in \eqref{eq:auxiliary_function_g}.
We have verified that the first two moments are increasing functions of $\mu$ (and hence decrease with
$\alpha$). Similarly higher moments can be computed using Mathematica, but we do not
write them explicitly as they have rather long expressions.

\begin{figure}
    \includegraphics[width=0.7\linewidth]{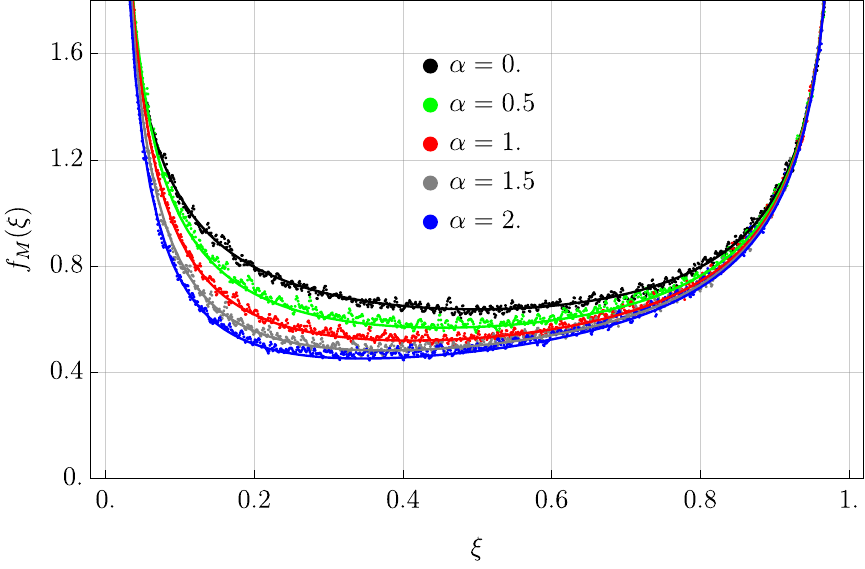}
    \caption{(Solid lines) Probability distribution $f_{\rm M}(\xi)$ computed analytically in \eqref{eq:f_M} for
different values of $\alpha$. (Points) Numerically obtained scaling function $f_M(\xi)$.
Simulation details are same as in Fig. \ref{fig:f_plus}.}
\label{fig:f_M}
\end{figure}

To plot the scaling function $f_M(\xi)$ and compare it with numerical simulations, we need to however
invert the Stieltjes transform \eqref{eq:stieltjes_time_of_max}. Fortunately, this can be done, at least formally,
by applying the Sokhotski–Plemelj formula~\cite{sokhotski-plamelij-wiki} which states that if the Stieltjes transform of any function $f(\xi)$ is given by
\begin{equation}
\int_0^1 \dd \xi\, \frac{f(\xi)}{z-\xi}= h(z)
\label{ST.1}
\end{equation}
for any complex $z$ excluding the cut $[0,1]$ on the real axis, then $f(\xi)$ for $\xi\in [0,1]$ on the
real axis can be expressed as
\begin{equation}
f(\xi)= \frac{1}{\pi}\, {\rm Im}\left[h(\xi-i0^+)\right]\, , \quad {\rm for}\quad 0\le \xi\le 1\, .
\label{IST.1}
\end{equation}  
Inverting \eqref{eq:stieltjes_time_of_max} using this formula, we get
\begin{equation}
    f_{M}(\xi) = \frac{\mc{N}}{\xi\, \pi}\, \Im\int_0^{\infty} \frac{\dd x}
{\left[g_\mu\left(-\ii\sqrt{\frac{1}{\xi}-1}\, x\right)\right]}\, \frac{ g'_\mu(x)}{g_\mu(x)}\, , 
\label{eq:f_M}
\end{equation}
where $\mc{N}$ and $g_\mu(x)$ are given respectively in Eqs. (\ref{eq:normalisation_M}) and 
(\ref{eq:auxiliary_function_g}). 
Exploiting the fact that the only singularities of the integrand are along the purely 
imaginary axis of the complex plane (a non-trivial fact that we have verified numerically), a 
wedge-shaped contour makes the integration easy. We have found empirically that a wedge 
with an angle $\omega \approx \pi /4$ with the real axis achieves the best accuracy (eee 
Section \ref{sec:calculation_f_M} and Fig. \ref{fig:complex_contour} for details).
Comparison of the numerical integration in formula \eqref{eq:f_M} and the sampling of $t_M$ 
from microscopic trajectories generated from the rules \eqref{eq:sluggish_rules} can be seen 
in Fig. \ref{fig:f_M}. 

The scaling function $f_M(\xi)$ is asymmetric around $\xi=1/2$ and is convex in $\xi\in [0,1]$, diverging
at the two edges $\xi\to 0$ and $\xi\to 1$. However, these asymptotic behaviors near the two edges
are not so easy to derive as in the cases of $f_+(\xi)$ and $f_{\rm l}(\xi)$. They can nevertheless be extracted
from \eqref{eq:f_M} exactly and we find
\begin{eqnarray}
\label{fM_asymp}
f_M(\xi) \approx \begin{cases}
& A_\mu\, \xi^{\mu-1}\quad , \quad\quad\quad \xi\to0^+ \\
\\
& B_\mu\, (1-\xi)^{-\frac{1}{2}}\quad , \quad \xi\to 1^-\quad\, ,
\end{cases}
\end{eqnarray}
where the two prefactors $A_\mu$ and $B_\mu$ are given explicitly by
\begin{eqnarray}
    A_\mu &= & \frac{2\mu\, 2^{-\mu}}{\Gamma(\mu)\,\Gamma(1-\mu)\,\Gamma(1+\mu)}\, 
\int_0^{\infty} \frac{dx\, x^{2\mu-1}}{g_\mu(x)}\, ,
\label{Amu.1} \\
    B_\mu & = &\frac{2^\mu}{\Gamma(1-\mu)\,\pi}\, \int_0^{\infty}\frac{\dd x}{g_\mu(x)}\, . 
\label{Bmu.1}
\end{eqnarray}
For the Brownian case $\mu=1/2$, one recovers $A_{1/2}=B_{1/2}= 1/\pi$.

Interestingly the inverse square root divergence as $\xi\to 1$ in Eq. (\ref{fM_asymp}) is universal, i.e.,
independent of $\mu=1/(\alpha+2)$.
A more important observation is that the all the three PDF's $f_+(\xi)$, $f_{\rm l}(\xi)$
and $f_M(\xi)$ have the same power law divergence $f(\xi)\sim \xi^{\theta-1}$ 
as $\xi \to 0^+$ where $\theta=\mu=1/(\alpha+2)$ is the persistence
exponent defined in \eqref{surv.1}.

\section{Derivation of the results}
\label{section:3}

In this Section we proceed with the derivation of the main results and it thus
represents the main technical part of the paper. The 
order of the derivation of the results is the same as the presentation in Section \ref{section:2}.
We start with $t_+$, follow it up by $t_{\rm l}$  and 
finally move to $t_{M}$. Out of the three, the derivation of the distribution of $t_M$ 
turns out to be the most challenging and nontrivial.

\subsection{Distribution of the occupation time $t_+$}
\label{sec:calculation_f_plus}

Here we derive the distribution of the occupation time $t_+$. The derivation consists of three steps.
We start with a backward Feynamn-Kac equation 
satisfied by the generating function by generalizing a technique used in \cite{Majumdar2005} to compute the
distribution of $t_+$ for a Brownian motion. This will give access to the double 
Laplace transform of $p_+(t|T)$ with respect to $t$ and $T$.
In the second step, we relate the double Laplace transform of $p_+(t|T)$ to the Stieltjes transform 
of the distribution assuming the scaling form \eqref{eq:p_plus_scaling}.
Finally, we invert the Stieltjes transform and find the distribution function $p_+(t|T)$ explicitly.

\subsubsection{Backward Feynman-Kac equation for the Laplace transform}

To derive the distribution of $t_+$ in \eqref{eq:t_plus_def} we first consider 
$p_+(t|T,x_0)$, i.e., the distribution of $t_+$, given the total duration $T$ and an arbitrary
initial position $x_0$. The reason for using an arbitrary $x_0$ is that we will use $x_0$
as a variable to derive a backward Feynamn-Kac equation. At the end of the calculation, we 
will set $x_0=0$ and compute $p_+(t|T)= p_+(t|T,x_0=0)$. It is convenient to introduce 
the Laplace transform of $p_+(t|T,x_0)$ with respect to $t$, i.e.,  
\begin{equation}\label{eq:first_laplace_transform}
\phi(x_0,T|p) = \int_0^{\infty}\dd t\, e^{-p\, t}\,p_+(t|T,x_0) = \braket{e^{-p\, t_+}}_{x_0}= 
\braket{ e^{-p\, \int_0^T
\Theta\left(x(t)\right)\, dt }}_{x_0} 
\end{equation}
where the average is taken over all trajectories starting at fixed $x(0) = x_0$. 
We now follow the backward Feynman-Kac approach used in Ref.~\cite{Majumdar2005} for the Brownian motion
to derive a partial differential equation for $\phi(x_0,T|p)$. The main idea is to consider the
stochastic moves during the small time interval $[0, dT]$ and write down the evolution equation
for $\phi(x_0,T|p)$ by taking the limit $dT\to 0$. Given the backward generator in Eq. (\ref{eq:backward_survivial}), it
is then straightforward to show that $\phi(x_0,T|p)$ evolves via (see Appendix A for the detailed derivation)
\begin{equation}\label{eq:backward_equation}
    \p_T \phi(x_0,T|p) =\frac{1}{|x_0|^\alpha} \p_{x_0}^2 \phi(x_0,T|p) - p \Theta(x_0)\phi(x_0,T|p)\, , 
\end{equation}
starting from the initial condition $\phi(x_0,T=0|p)=1$ that follows from the definition
in \eqref{eq:first_laplace_transform}.

To reduce the partial differential equation \eqref{eq:backward_equation} into an ordinary differential equation (ODE) in $x_0$,
it is further convenient to consider a second Laplace transform with respect to $T$ by defining
\begin{equation}\label{eq:second_laplace_transform}
    \tilde \phi(x_0|s, p) = \int_0^{\infty}\dd T \, e^{-s\, T}\,\phi(x_0,T|p)\, .
\end{equation}
Taking Laplace transform of \eqref{eq:backward_equation} with respect to $T$ and using the initial condition
 $\phi(x_0,T=0|p)=1$, we get our desired second order ODE
\begin{equation}\label{eq:laplace_transformed_eq}
    \frac{1}{|x_0|^\alpha}\frac{d^2 \tilde \phi(x_0|s,p)}{dx_0^2} - \left[p\, \theta(x_0)+ s\right]\, 
\tilde \phi(x_0|s,p) = -1 \, .
\end{equation}
The boundary conditions are fixed by noticing that for $x_0\to -\infty$ the walker can not cross the origin and
can not become positive for any finite $T$ and consequently $t_+=0$ from its definition \eqref{eq:t_plus_def}. This implies, 
from \eqref{eq:first_laplace_transform} and \eqref{eq:second_laplace_transform}, that $\tilde \phi(x_0\to -\infty|s,p)=\frac{1}{s}$. On the other hand if 
$x_0\to +\infty$ the walker always stays positive so that $t_+=T$ implying, 
again from \eqref{eq:first_laplace_transform} and \eqref{eq:second_laplace_transform}, $\tilde \phi(x_0\to +\infty|s,p) = 
\frac{1}{p+s}$. We also have to impose the matching condition at $x_0=0$,
i.e., to ensure the continuity of the solution and its derivative at the origin $x_0=0$.

The solution of this second order non-homogeneous equation \eqref{eq:laplace_transformed_eq} can be readily found
in the two regions $x_0\ge 0$ and $x_0\le 0$ separately by first making an appropriate constant shift in each region to make
the equation homogeneous and then solving the resulting homogeneous ODE. We get the general solution as
\begin{eqnarray}\label{eq:solution_phi_preliminary}
\tilde \phi(x_0|s,p) = \begin{cases} 
& c_1\, \sqrt{x_0}\, I_{\mu}\left(2\mu(s + p)^{\frac{1}{2}}x_0^{\frac{1}{2\mu}}\right) + c_2\, \sqrt{x_0}\, 
K_{\mu}\left(2\mu(s + p)^{\frac{1}{2}}x_0^{\frac{1}{2\mu}}\right) + \frac{1}{s+p} \quad {\rm for}\quad  x_0\geq 0 \\
&\\
& c_3\, \sqrt{-x_0}\, I_{\mu}\left(2\mu\,s^{\frac{1}{2}}(-x_0)^{\frac{1}{2\mu}}\right) + c_4\, \sqrt{-x_0}\, 
K_{\mu}\left(2\mu\,s^{\frac{1}{2}}(-x_0)^{\frac{1}{2\mu}}\right)+\frac{1}{s} \quad {\rm for}\quad  x_0<0\, ,
    \end{cases}
\end{eqnarray}
where $c_1$, $c_2$, $c_3$ and $c_4$ are arbitrary constants that can be fixed from the two boundary condition
as $x_0\to \pm\infty$ and the two matching conditions at $x_0=0$.

For large argument we use the following asymptotic behaviors of the Bessel function~\cite{watson_treatise_bessel}
\begin{equation}\label{eq:besself_first_large_argument}
    I_\mu(z) \sim \frac{e^z}{(2\pi z)^{1/2}}\quad , \quad z\to +\infty
\end{equation}
and
\begin{equation}\label{eq:besself_second_large_argument}
    K_\mu(z) \sim \left(\frac{\pi}{2z}\right)^{1/2}e^{-z}\quad , \quad z\to +\infty
\end{equation}
to fix $c_1=c_3=0$ in order to have a non-diverging solution as $x_0\to \infty$ . This gives
\begin{equation}\label{eq:solution_phi_preliminary2}
    \tilde \phi(x_0|s,p) = \begin{cases}
         c_2\sqrt{x_0}\, K_{\mu}\left(2\mu(s + p)^{\frac{1}{2}}x_0^{\frac{1}{2\mu}}\right) + 
\frac{1}{s+p\, }\quad,\quad & x_0\geq 0
        \\
        c_4\sqrt{-x_0}\, K_{\mu}\left(2\mu\,s^{\frac{1}{2}}(-x_0)^{\frac{1}{2\mu}}\right)+
\frac{1}{s} \quad,\quad & 
x_0<0
    \end{cases}\quad .
\end{equation}
To impose the matching conditions at $x_0=0$ to fix the remaining constants $c_2$ and $c_4$,
we use the following small argument behaviors~\cite{watson_treatise_bessel}
\begin{eqnarray}
    I_\mu(z) &= & \frac{\left(\frac{z}{2}\right)^\mu}{ \Gamma(1+\mu)} + \dots\quad \quad\quad\quad\quad\quad\quad \quad z \to 0
\label{eq:bessel_first_small_argument} \\
\\
    K_\mu(z) &= & \frac{\Gamma(\mu)}{2\left(\frac{z}{2}\right)^\mu} - 
\left(\frac{z}{2}\right)^\mu \frac{\Gamma(1-\mu)}{2\mu}+\dots\quad  \quad z\to 0\quad .
\label{eq:bessel_second_small_argument}
\end{eqnarray}
From the continuity condition at $x_0=0$ we get
\begin{equation}\label{eq:function_continuity}    
\frac{c_2}{2}\, \Gamma\left(\mu\right)\left[\mu(s+p)^{1/2}\right]^{-\mu} + \frac{1}{s + p}= \frac{c_4}{2}\, 
\Gamma\left(\mu\right)\left[\mu\,s^{1/2}\right]^{-\mu}+\frac{1}{s}\quad ,
\end{equation}
while from the continuity of the first derivative at $x_0=0$ we get
\begin{equation}\label{eq:derivative_continuity}
    -\frac{c_2}{2}\, \frac{\Gamma\left(1-\mu\right)}{\mu}(s + p)^{\frac{\mu}{2}} = \frac{c_4}{2}\, 
\frac{\Gamma\left(1-\mu\right)}{\mu}s^{\frac{\mu}{2}}\quad .
\end{equation}
Solving these two linear equations, we get $c_2$ and $c_4$ explicitly
\begin{equation}
    c_2 = \frac{2 p \mu^\mu s^{\mu -1}
   (s +p)^{\frac{\mu}{2}-1}}{s\, \Gamma
   \left(\mu\right) \left(s
   ^{\mu}+(s +p)^{\mu}\right)}\quad\quad, \quad \quad
    c_4 = -\frac{2 \mu^{\mu} p s ^{\frac{\mu}{2}-1}
   (s +p)^{\mu-1}}{\Gamma
   \left(\mu\right) \left(s
   ^{\mu}+(s +p)^{\mu}\right)}\quad .
\end{equation}
Substituting $c_2$ and $c_4$ in Eq. (\ref{eq:solution_phi_preliminary2}) then gives the full solution
$\tilde\phi (x_0|s,p)$ for arbitrary $x_0$. The result simplifies a lot by setting $x_0=0$.
Using the asymptotic behavior
\begin{equation}\label{eq:limit_modified_bessel_second}
    \lim_{x_0\to 0^+}\sqrt{x_0}\, K_{\mu}\left(2\mu s^{1/2}x_0^{\frac{1}{2\mu}}\right)=
\frac{1}{2}\, \frac{s^{-\frac{\mu}{2}}}{\mu^\mu}\,  \Gamma \left(\mu\right)\quad .
\end{equation}
in \eqref{eq:solution_phi_preliminary2}), we finally get 
\begin{equation}\label{eq:double_laplace_preliminary}
    \tilde \phi(x_0=0|s,p) = \int_0^{\infty} dT\, e^{-s\, T} \int_0^T dt\, e^{-p\, t}\, p_+(t|T)= 
\frac{1}{s}\left[1- \frac{p(s+p)^{\mu-1}}{s
   ^{\mu}+(s +p)^{\mu}}\right]\quad .
\end{equation}
For normal diffusion, i.e., for $\alpha=0$ (or equivalently $\mu=1/2$), one recovers the
Brownian result $\tilde \phi(x_0=0|s,p)=1/\sqrt{s (s+p)}$~\cite{Majumdar2005}.

\subsubsection{Stieltjes transform}

Up to now we have computed the double Laplace transform of the distribution of $t_+$ in \eqref{eq:double_laplace_preliminary}. 
To extract the scaling function $f_+(\xi)$ by inverting the double Laplace transform is not easy. For this,
we first transform the double Laplace transform in \eqref{eq:double_laplace_preliminary} to a more familiar
single variable transform known as the Stieltjes transform. This is achieved easily by substituting the
anticipated scaling form $p_+(t|T)= (1/T)\, f_+(t/T)$ 
(expected to hold for all $t_+$ and $T$ in the continuous space time model)
in \eqref{eq:double_laplace_preliminary}. This gives
\begin{equation}
\label{eq:stieltjes_plus}
    \tilde \phi(x_0=0|s,p)=\int_0^{+\infty}\dd T \, e^{-s T}\int_0^T \dd t \, e^{-p t} \frac{1}{T}f_+(t/T)
    = \int_0^1 \dd \xi \frac{f_+(\xi)}{s + p \xi}
\end{equation}
where in the last line we changed variables $t=\xi T$ and exchanged the integrals. 
This almost has a form close to Stieltjes transform , but not quite yet. To proceed further,
let us rewrite the right hand side (r.h.s) of \eqref{eq:double_laplace_preliminary} in a
more symmetric manner
\begin{equation}\label{double_laplace_preliminary2}
    \tilde \phi(x_0=0|s,p) = \frac{s^{\mu-1}}{s^{\mu} + (s + p)^{\mu}} +  
\frac{(s+p)^{\mu-1}}{s^{\mu} + (s + p)^{\mu}}\quad .
\end{equation}
Next we rewrite
\begin{equation}\label{eq:f_plus_splitting}
    f_+(\xi) = y_1(\xi) + y_2(1-\xi)
\end{equation}
such that
\begin{equation}
    \int_0^1 \dd \xi \frac{y_1(\xi)}{\alpha + p \xi} =\frac{s^{\mu-1}}{s^{\mu} + (s + p)^{\mu}} \, \quad
{\rm and}\quad 
    \int_0^1 \dd \xi \frac{y_2(1-\xi)}{s + p \xi} =\frac{(s+p)^{\mu-1}}{s^{\mu} + (s + p)^{\mu}}\quad .
\end{equation}
Changing variables $1-\xi \mapsto \xi$ and sending $s \mapsto s -p$ and $p\mapsto -p$ proves that
\begin{equation}\label{eq:y1_equal_y2}
    y_1(\xi)=y_2(\xi)\, ,
\end{equation}
so we can focus on $y_1(\xi)$ only.

Scaling $\alpha = -p z$ we obtain
\begin{equation}\label{eq:R_plut_definition}
    \int_0^1 \dd x \frac{y_1(\xi)}{z-\xi} = 
\frac{1}{z}\frac{1}{\left[1+\left(1-\frac{1}{z}\right)^{\mu}\right]}
\equiv R_+(z)\quad.
\end{equation}
Using \eqref{eq:f_plus_splitting}) and the result \eqref{eq:R_plut_definition}, we then arrive at the result
announced in \eqref{eq:stieltjes_f_plus_result}.
Below we show how to invert this single variable Stieltjes transform.

\subsubsection{Inversion and exact calculation of $f_+(\xi)$.}

The inversion of the Stieltjes transform is done by promoting $z\in \mathbb{C}$ and by applying 
the Sokhotski-Plemelj formula stated in Eq. (\ref{IST.1})
\begin{equation}\label{eq:SP_formula}
    y_1(\xi) = \frac{1}{\pi}\, \Im R_+(\xi-\ii 0^+)
\end{equation}
where $R_+(z)$ is defined in \eqref{eq:R_plut_definition} and $\xi\in\mathbb{R}$. Setting
\begin{equation}
    1-\frac{1}{\xi - \ii \epsilon} = \rho(\xi) \, e^{\ii \theta(\xi)}
\end{equation}
where the modulus $\rho(\xi)$ and the phase $\theta(\xi)$ 
of the left hand side (l.h.s) are given by 
\begin{equation}\label{mod_phase}
    \rho(\xi) = \sqrt{\left(1-\frac{\xi}{\xi^2+\epsilon^2}\right)^2 + \frac{\epsilon^2}{\xi^2 + \epsilon^2}}\, \quad {\rm and}
\quad 
    \theta(\xi) = \arg\left[1-\frac{1}{\xi-\ii\epsilon}\right]\quad .
\end{equation}
We also need the limits
\begin{equation}\label{eq:limit_rho}
    \lim_{\epsilon\to0^+} \rho(\xi) = \frac{1-\xi}{\xi}\,, \quad 0<\xi<1 \, \quad {\rm and}\quad
    \lim_{\epsilon\to 0^+}\theta(\xi) = \pi\, , \quad 0<\xi<1\quad .
\end{equation}
Using these results we can readily compute the limit in \eqref{eq:SP_formula} (we use below the shorthand
notation $\rho(\xi)\equiv \rho$ and $\theta(\xi)\equiv \theta$ for convenience)
\begin{eqnarray}
    \frac{1}{\pi}\lim_{\epsilon\to 0^+}\Im R_+(\xi-\ii \epsilon)
    &= &-  \frac{1}{\pi}\lim_{\epsilon\to 0^+}\Im\left\{\frac{1}{\xi-\ii 
\epsilon}\frac{1}{1 + \rho^{\mu}\left[\cos\left(\mu \theta\right) + \ii \sin\left(\mu \theta\right)\right]}\right\} \nonumber \\
    & = & - \frac{1}{\pi}\lim_{\epsilon\to 0^+}\left\{\frac{\xi+\ii \epsilon}{\xi^2+\epsilon^2}\frac{1+
\rho^{\mu}\cos\left(\mu\theta\right) 
+ - \ii \rho^{\mu \theta}\sin\left(\mu\theta\right)}{\left(1 + \rho^{\mu \theta}\cos\left(\mu\theta\right)\right)^2 
+  \rho^{2\mu \theta}\left(\sin\left(\mu\theta\right)^2\right)} \right\}
    \nonumber \\
    & = & \frac{1}{\pi}\lim_{\epsilon\to 0^+}\left\{\frac{\xi}{\xi^2 + \epsilon^2}\frac{\rho^{\mu 
\theta}\sin\left(\mu\theta\right)}{\left(1 + \rho^{\mu \theta}\cos\left(\mu\theta\right)\right)^2 +  
\rho^{2\mu \theta}\left(\sin\left(\mu\theta\right)^2\right)}\right\}\quad .
\end{eqnarray}

Using the two limits in \eqref{eq:limit_rho} in the last line and rearranging the terms we obtain, 
from the Sokhotski-Plemelj formula \eqref{eq:SP_formula}, for $\xi\in(0,1)$,
\begin{align}
    y_1(\xi) = \frac{\sin\left(\pi \mu\right)}{\pi}\frac{\xi^{\mu - 1}(1-\xi)^{\mu}}{\xi^{2\mu} + 2\left[\xi(1-\xi)\right]^{\mu}\cos\left(\pi \mu\right) + \left(1-\xi\right)^{2\mu}}\quad .
\end{align}
Then using \eqref{eq:y1_equal_y2} and \eqref{eq:f_plus_splitting} and after some simple algebra 
we arrive at the final result \eqref{eq:f_plus} of Section \ref{section:2}.

\subsection{Distribution of the last passage time $t_{\rm l}$}
\label{sec:calculation_f_l}

\begin{figure}
    \centering
    \includegraphics[width=0.7\linewidth]{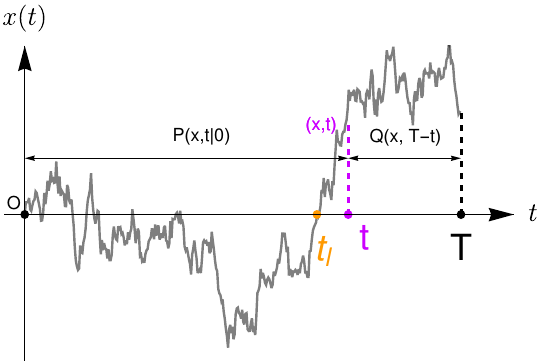}
    \caption{Schematic representation of the path decomposition behind the 
calculation of the distribution of $t_{\rm l}$ in \eqref{eq:distribution_t_last_decomposition}. 
The contribution of the left path is provided by the free propagator $P(x,t|x_0=0)$ 
in \eqref{eq:propagator_free} while that to the right part comes from the survival probability 
$Q(x, T-t)$ in \eqref{eq:survival_free}.}
    \label{fig:path_decomposition_last}
\end{figure}

The distribution of $t_{\rm l}$ turns out to be the easiest to derive and can be computed 
directly from the the knownledge of the propagator and the survival probability derived in Section \ref{section:2}. 
This follows from the fact that the cumulative distribution of $t_{\rm l}$, for a sluggish walker starting
at the origin, can be simply expressed as
\begin{equation}\label{eq:distribution_t_last_decomposition}
    \Prob\left[t_{\rm l} \leq t|T\right] = \int_{-\infty}^{+\infty}\dd x \, P(x,t|0,0)\, Q(x, T-t)\quad ,
\end{equation}
where $P(x,t|x_0,0)$ is the propagator from $x_0$ to $x$ in time $t$ and $Q(x,t)$ is the survival probability 
starting at $x$. The reasoning behind \eqref{eq:distribution_t_last_decomposition} is as follows. Consider the event when the
last crossing of the origin occurs at time $t_l\le t$. Then the whole trajectory can be decomposed
into two time intervals: (i) $[0,t]$ and (ii) $[t, T]$. In the the first interval we have a path going from the 
starting position $x_0$ at time $t=0$ to some position $x$ at time $t\in[0,T]$ with the associated weight 
given by the propagator $P(x,t|x_0,0)$. In the second interval $[t,T]$, the path can not cross the origin
since the last passage through the origin occurred before $t$. Hence, during $[t,T]$ the path has to 
propagate from $x$ to
any final position at $T$ without crossing the origin. The associated weight is precisely the survival probability
$Q(x,T-t)$ during the interval $(T-t)$, starting at $x$. This path decomposition is explained schematically
in Fig. \ref{fig:path_decomposition_last}.

From the cumulative distribution \eqref{eq:distribution_t_last_decomposition}, the PDF of $t_l$ can be
simply obtained by diffferentiating with respect to $t$, leading to
\begin{equation}\label{eq:f_last_density}
p_{\rm l}(t|T) =  \p_t \int_{-\infty}^{+\infty}\dd x \, P(x,t|x_0=0,0)\, Q(x, T-t) \quad .  
\end{equation}
Substituting the scaling forms, $P(x,t|0,0)= t^{-\mu}\, H_P(x t^{\mu})$ and $Q(x,t)= H_Q(x\, t^{-\mu})$ respectively
from Eqs. (\ref{eq:scaling_ansatz_propagator}) and (\ref{eq:scaling_ansatz_survival}), in 
Eq. (\ref{eq:f_last_density}) and rescaling $x\, t^{-\mu}\to x$, one gets 
\begin{equation}
\label{pltT.1}
p_{\rm l}(t|T) =  2\, \int_{0}^{+\infty}\dd x \, H_P\left(x\right)\, 
\p_t H_Q\left(\frac{x}{\left(\frac{T}{t}-1\right)^{\mu}}\right)\, .
\end{equation}
The factor $2$ comes from the $z\to -z$ symmetry of the scaling functions $H_P(z)$ and $H_Q(z)$.
Using the explicit form of $H_Q(z)$ from Eq. (\ref{eq:survival_free}) and performing the time 
derivative in \eqref{pltT.1} leads,
after straightforward algebra, to the result
\begin{equation}
\label{pltT_final}
p_{\rm l}(t|T)= \frac{1}{T} \frac{\sin(\pi\mu)}{\pi}\frac{1}{(t/T)^{1-\mu}(1-t/T)^\mu}\, ,
\end{equation}
which was announced in \eqref{eq:p_l} and \eqref{eq:f_l}.

\subsection{Distribution of the time $t_M$ to reach the maximum}
\label{sec:calculation_f_M}

In this subsection we derive the probability distribution $p_M(t_M|T)$ of the
time instant $t_M$ in the
interval $[0,T]$ at which the sluggish walker is maxmally displaced on the positive side of the origin.
We note that while $t_M$ is a random variable taking different values in different realizations, for
convenience we will slightly abuse the notations and denote by $t_M$ the value of this random variable.
In the hydrodynamic limit we expect this distribution to have the scaling form
\begin{equation}
p_M(t_M|T)= \frac{1}{T}\, f_M\left(\frac{t_M}{T}\right)\, ,
\label{pm_scale.1}
\end{equation}
and our goal in this subsection is to compute this scaling function $f_M(\xi)$. 
To compute $p_M(t_M|T)$, it turns out to be convenient to first compute the joint distribution
$P(M,t_M|T)$ of the actual value $M$ of the maximum and its time of occurrence $t_M$. This
joint distribution is normalised to unity, i.e.,   
\begin{equation}
\label{eq:normalisation_joint}
\int_0^{\infty} dM\, \int_0^T dt\, P(M,t|T) =1\, .
\end{equation}
Knowing this joint distribution, one can then marginalize $M$ to obtain the PDF $p_M(t_M|T)$
\begin{equation}
p_M(t_M|T)= \int_0^{\infty} dM\, P(M,t_M|T)\, .
\label{marginal.1}
\end{equation}

To compute the joint distribution $P(M,t_M|T)$, we can use a path decomposition technique that exploits
again the Markov property of the process. This is best explained with the help of
Fig. ({\ref{fig:path_decomposition}). Consider a typical trajectory of the process where
the maximum occurs at $t_M$ with value $M$. We first divide the interval $[0,T]$ into
two parts: (i) the `left' part over $[0,t_M]$ and (ii) the `right' part over $[t_M,T]$.
In the left part, the trajectory, starting from $0$, has to remain below $M$ till time $t_M$
and it arrives at its maximal value $M$ exactly at the instant $t_M$.
On the right side, the process, starting at $M$ exactly at $t_M$,
has to stay below $M$ over the interval $[t_M,T]$. These constraints are necessary to
ensure that the maximum occurs at $t_M$ with value $M$. The constraint of staying below the level $M$
is normally implemented by imposing an absorbing boundary condition at the level $M$.
However, for a continuous time process such as the Brownian motion or its cousins, it is impossible
to implement the two constraints simultaneously, namely that the process stays below $M$ during $[0,t_M]$ and
$[t_M,T]$ and that it also arrives exactly at $M$ at $t$. This is because if a continuous time process
such as the Brownian motion
reaches a level $M$ at some time $t_M$, it must cross and recross the same level infinitely often immediately
before or after $t_M$.
To circumvent this technical difficulty, one can introduce
a small cut-off $\epsilon$~\cite{MC2005}, i.e., we impose that the trajectory arrives at $M-\epsilon$ (instead of $M$) at $t_M$,
and stays below $M$ during the full interval $[0,T]$. Then, there is no problem in imposing an
absorbing boundary condition at $M$. At the very end of the calculation, one needs to take
the $\epsilon\to 0$ limit in an appropriate way. This $\epsilon$-path decomposition method was
introduced in Ref.~\cite{MRKY2008} in the context of computing the distribution of $t_M$
for constrained Brownian motions such as excursions and meanders and since then, has been used
in numerous other contexts~\cite{Schehr2010,MMSS2021,MMS2019,MMS2020,MMS2021,MMS2022,PCMS13,PCMS15}.
Here we adopt this $\epsilon$-path decomposition technique to the sluggish random walk process and express
the joint distribution as the product of the probability weights of the left part $[0,t_M]$ and the
right part $[t_M,T]$ as
\begin{equation}\label{eq:joint_distribution_M_t_M}
P(M,t_M|T) \propto G_M(M-\epsilon, t_M|0)\, Q(M-\epsilon, T-t_M|M)\, .
\end{equation}
Here $G_M(x,t|x_0)$ is the probability density that the walker arrives at $x$ at time $t$ starting form $x_0$,
and is constrained to stay below the level $M$ during the interval $[0,t]$.
The quantity $Q(x_0,t|M)$ is the survival probability that the walker, starting at $x_0<M$, stays below
the level $M$ up to time $t$. The proportionality constant in \eqref{eq:joint_distribution_M_t_M}
gets fixed from the normalization condition \eqref{eq:normalisation_joint}.
Below we compute $Q(M-\epsilon, T-t_M|M)$ and $ G_M(M-\epsilon, t_M|0)$ separately, starting with the former one since it is
simpler.

\begin{figure}
    \centering
    \includegraphics[width=0.7\linewidth]{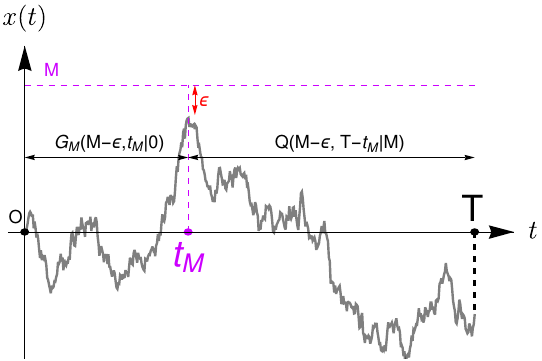}
 \caption{Decomposition of a generic sample path of the sluggish walker process, starting at the origin,
during $[0,T]$ into the left part over interval $[0,t_M]$ and the right part over the time interval $[t_M,T]$ 
respectively. On the left part the walker, starting from the origin, arrives at $M-\epsilon$
at time $t_M$ while staying below the level $M$ during $[0,t_M]$. The probability weight associated to
the left part is $G_M(M-\epsilon, t_M|0)$. On the right part, the process, starting at $M-\epsilon$ at $t_M$
propagates up to $T$ while staying below the level $M$ during $[t_M,T]$. The probability weight
associated to this right part is $Q(M-\epsilon, T-t_M|M)$. Taking the product as in Eq. (\ref{eq:joint_distribution_M_t_M})
gives the total probability weight of the path over $[0,T]$.}
\label{fig:path_decomposition}
\end{figure}

\subsubsection{Survival probability with absorbing boundary at $M$}

Here we want to compute the survival probability $Q(x_0,t|M)$ appearing in \eqref{eq:joint_distribution_M_t_M}.
This denotes the probability that the sluggish walker, starting at $x_0\le M$ at $t=0$, stays below the level $M$
up to time $t$. Incidentally, $Q(x_0,t|M)$ also has another probabilistic interpretation. Let $M(t)$ denote
the maximum of the process up to time $t$, starting from $x_0$. Then $M(t)$ is a random variable that fluctuates from sample to sample
and computing its distribution is an important extreme value problem~\cite{Majumdar2020,Majumdar2024}.
It is easy to see that the cumulative distribution of $M(t)$ precisely coincides with the survival probability 
\begin{equation}
{\rm Prob.}\left[M(t)\le M\right]= Q(x_0,t|M)\, .
\label{max_cumul.1}
\end{equation}
This is because ${\rm Prob.}\left[M(t)\le M\right]$ counts all the events where $M(t)$ stays below $M$ up to time $t$ which, by definition,
is simply $Q(x_0,t|M)$. The probability $Q(x_0,t|M)$ satisfies the same backward equation \eqref{eq:backward_survivial} in the
regime $x_0\le M$, but now with the boundary conditions 
 \begin{equation}\label{eq:survival_boundary_conditions}
     Q(x_0=M,t|M) = 0\, ,  \quad {\rm and}\quad  Q(x_0\to -\infty,t|M)=1\, , 
\end{equation}
and the initial condition
\begin{equation}
Q(x_0,t=0|M)=1 \, \quad {\rm for}\quad x_0<M\, .
\label{initial_Q}
\end{equation}
The first boundary condition in \eqref{eq:survival_boundary_conditions} imposes the trajectory to stay below $M$ during $t$,
while the second one is due to the fact that the trajectory always survives (with probability one) up to a finite time $t$
if it starts very far away from the absorbing boundary at $M$. 

Next we define the Laplace transform
\begin{equation}
\tilde Q(x_0,s|M)= \int_0^{\infty} dt\, e^{-s\, t}\, Q(x_0,t|M)\, . 
\label{laplace_def.1}
\end{equation}
Taking the Laplace transform of \eqref{eq:backward_survivial} with respect to $t$ and using the initial condition
\eqref{initial_Q}, the partial differential equation \eqref{eq:backward_survivial} reduces to an ODE in $x_0\le M$
 \begin{equation}
\label{survival_ODE.1}
    \frac{1}{|x_0|^\alpha}\, \frac{d^2\tilde  Q(x_0,s|M)}{dx_0^2} -s \tilde Q(x_0,s|M) =-1
 \end{equation}
with the boundary conditions: $\tilde Q(x_0=M,s|M)=0$ and $\tilde Q(x_0\to -\infty,s|M)=1/s$.
To solve this ODE, we first note that due to the presence of $|x_0|^{-\alpha}$ term in \eqref{survival_ODE.1},
we need to solve in two different regions $0\le x_0\le M$ and $x_0\le 0$ and then match the solution
and its derivative at $x_0=0$. These solutions in the two regions can be readily found to be
\begin{align}\label{eq:survival_probability}
    \tilde Q(x_0,s) = 
    \begin{cases}
        b_1\, \sqrt{x_0}\, I_{\mu}\left(2\mu s^{\frac{1}{2}}x_0^{\frac{1}{2\mu}}\right) + 
b_2\, \sqrt{x_0}\, K_{\mu}\left(2\mu s^{\frac{1}{2}}x_0^{\frac{1}{2\mu}}\right) + \frac{1}{s}\, ,& \quad 0\leq x_0\leq M
        \\
b_3\,\sqrt{-x_0}\, I_{\mu}\left(2\mu s^{\frac{1}{2}}(-x_0)^{{\frac{1}{2\mu}}}\right) +b_4\, \sqrt{-x_0}\, K_{\mu}\left(2\mu s^{\frac{1}{2}}(-x_0)^{{\frac{1}{2\mu}}}\right)+\frac{1}{s}\, ,  & \quad x_0<0
    \end{cases}
\end{align}
where $\mu = 1/(\alpha + 2)$ as usual and $b_i$'s are arbitrary constants to be fixed from the boundary conditions at $x_0=M$,
$x_0\to -\infty$ and the two matching conditions at $x_0=0$. Since the Bessel function $I_\mu(z)$ diverges
exponentially as $z\to \infty$ (see \eqref{eq:besself_first_large_argument}), we must have $b_3=0$ to satisfy
the boundary condition $\tilde Q(x_0\to -\infty,s|M)=1/s$.  

To find the remaining constants, we first impose the matching of the solution and its derivative at $x_0=0$
by using the small argument expansion of the Bessel functions in \eqref{eq:bessel_first_small_argument} 
and \eqref{eq:bessel_second_small_argument}. The continuity of the solution at $x_0=0$ gives the relation
\begin{equation}
      b_2\,\frac{1}{2} \frac{s^{-\mu / 2}}{\mu^\mu}\Gamma\left(\mu\right) + 
\frac{1}{s} = b_4\, \frac{1}{2} \frac{s^{-\mu / 2}}{\mu^\mu}\Gamma\left(\mu\right) + \frac{1}{s}
\end{equation}
which implies $b_2 = b_4$. The continuity of the derivative gives an additional relation 
\begin{equation}
    b_1\, \frac{\mu^\mu s^{\frac{\mu}{2}}}{\Gamma\left(1+\mu\right)}  - b_2\, 
\frac{\Gamma\left(1-\mu\right)}{2\mu}\mu^\mu s^{\frac{\mu}{2}} = b_4\,\frac{\Gamma\left(1-\mu\right)}{2\mu}\mu^\mu s^{\mu / 2}
\end{equation}
which implies
\begin{equation}\label{eq:c1}
    b_1 = b_2\frac{\Gamma\left(1+\mu\right)\Gamma\left(1-\mu\right)}{\mu} =  \frac{\pi}{\sin (\pi \mu)}b_2
\end{equation}
where in the last line we have used $\Gamma(1+\mu)/\mu = \Gamma(\mu)$ and the reflection formula 
$\Gamma(\mu) \Gamma(1-\mu) = \pi / \sin(\pi\mu)$.
Imposing now the remaining boundary condition $\tilde Q(x_0=M,s|M)=0$ we obtain
\begin{equation}\label{eq:c2}
    b_2 = -\frac{1}{s \sqrt{M}}\frac{1}{\frac{\pi}{\sin (\pi \mu)}I_{\mu}\left(2\mu s^{\frac{1}{2}}M^{\frac{1}{2\mu}}\right) + 
K_{\mu}\left(2 \mu s^{\frac{1}{2}}M^{\frac{1}{2\mu}}\right)}\quad .
\end{equation}
Using \eqref{eq:c1} and \eqref{eq:c2} in \eqref{eq:survival_probability} gives the
full solution 
\begin{equation}
\label{sol_lap.1}
\tilde Q(x_0,s|M) = \frac{1}{s}\left[1-\sqrt{\frac{x_0}{M}}\, 
\frac{\frac{\pi}{\sin (\pi \mu)} I_{\mu}\left(2\mu s^{\frac{1}{2}}x_0^{\frac{1}{2\mu}}\right) + 
K_\mu\left(2\mu s^{\frac{1}{2}}x_0^{\frac{1}{2\mu}}\right)}{\frac{\pi}{\sin (\pi \mu)}
I_{\mu}\left(2\mu s^{\frac{1}{2}}M^{\frac{1}{2\mu}}\right) + K_{\mu}\left(2 \mu s^{\frac{1}{2}}M^{\frac{1}{2\mu}}\right)}\right]\quad .
\end{equation}

Using the definition of the modified Bessel function of the second kind $K_\mu(x) = \frac{\pi}{2\sin(\pi \mu)}(I_{-\mu}(x)-I_\mu(x))$ 
the result can be expressed in a more symmetric, albeit equivalent, form
\begin{equation}\label{eq:exact_survival_probability}
    \tilde Q(x_0,s|M) = \frac{1}{s}\left[1-\sqrt{\frac{x_0}{M}}\frac{ I_{\mu}\left(2\mu s^{\frac{1}{2}}x_0^{\frac{1}{2\mu}}\right) + 
I_{-\mu}\left(2\mu s^{\frac{1}{2}}x_0^{\frac{1}{2\mu}}\right)}{I_{\mu}\left(2\mu s^{\frac{1}{2}}M^{\frac{1}{2\mu}}\right) 
+ I_{-\mu}\left(2 \mu s^{\frac{1}{2}}M^{\frac{1}{2\mu}}\right)}\right]
 = \frac{1}{s}\left[1-\frac{\sqrt{x_0}r_\mu(s^\mu x_0)}{\sqrt{M}r_\mu(s^\mu M)}\right]\, ,
\end{equation}
where in the last line we have introduced the function 
\begin{equation}\label{eq:r_function}
    r_\mu(z) =I_\mu\left(2\mu z^{\frac{1}{2\mu}}\right) + I_{-\mu}\left(2\mu z^{\frac{1}{2\mu}}\right)
\end{equation}
that makes the final expression a bit more compact. Note, in particular, that for the process starting at $x_0=0$, 
Eq. (\ref{eq:exact_survival_probability}) simplifies to
\begin{equation}
 \tilde Q(0,s|M) =\lim_{x_0\to 0} \tilde Q(x,x_0,s)
     = \frac{1}{s}\left[1 -  \frac{1}{\sqrt{M}r_\mu(s^\mu M)}\frac{1}{2} 
\frac{s^{-\frac{\mu}{2}}}{\mu^\mu}  \Gamma \left(\mu\right)\right]\quad .
\label{Q0.1}
\end{equation}
This result thus generalizes the one derived previously for the special case $\alpha=1$, i.e., 
for $\mu=1/3$~\cite{Zodage_2023}.

Having obtained the complete solution of the Laplace transform $\tilde Q(x_0,s|M)$ in \eqref{eq:exact_survival_probability}, we
now set $x_0=M-\epsilon$ and compute $\tilde Q(M-\epsilon,s|M)$ to leading order in small $\epsilon$, since we will need
precisely this object for the right part of the path probability in \eqref{eq:joint_distribution_M_t_M}. This gives
\begin{equation}
    \tilde Q(M-\epsilon,s|M) =\frac{1}{s}\left[1-\frac{\sqrt{M-\epsilon}r_\mu(s^\mu (M-\epsilon))}{\sqrt{M}r_\mu(s^\mu M)}\right]\, .
\end{equation}
Making a Taylor expansion in $\epsilon$ and keeping only the leading term gives
\begin{equation}\label{eq:survival_expanded}
    \tilde Q(M-\epsilon, s|M) = \frac{\epsilon}{s}\, \frac{\p_M\left[\sqrt{M}r_\mu\left(s^\mu M\right)\right]}{\sqrt{M}
r_\mu\left(s^\mu M\right)} + O(\epsilon^2)\quad .
\end{equation}
Finally, inverting the Laplace transform formally, we get the leading order behavior for the right part of the
probability weight in Eq. (\ref{eq:joint_distribution_M_t_M})
\begin{equation}
Q(M-\epsilon, T-t_M|M)= \frac{\epsilon}{\sqrt{M}}\, {\cal L}^{-1}_{s\to (T-t_M)}\left[ \frac{\p_M\left[\sqrt{M}r_\mu\left(s^\mu M\right)\right]}
{s\, r_\mu\left(s^\mu M\right)}\right] + O(\epsilon^2)\, ,
\label{linv_right}
\end{equation}  
where ${\cal L}^{-1}_{s\to t}$ indicates the inverse Laplace transform with respect to $s$ and $t$ denotes the conjugate of $s$.

\subsubsection{The propagator with an absorbing boundary}

We now turn to the left part of the probability weight in \eqref{eq:joint_distribution_M_t_M}.
For this, we need to compute the constrained propagator $G_M(x,t|x_0)$ denoting the probability density
to arrive at $x$, starting at $x_0$ and staying below the level $M$ in $[0,t]$. Clearly, $G_M(x,t|x_0)$
also satisfies the same forward Fokker-Planck equation \eqref{eq:FPE} as the free propagator, i.e.,
\begin{equation}
 \frac{\partial G_M(x,t|x_0)}{\partial t} = \frac{\partial^2}{\partial x^2} \left[\frac{1}{|x|^\alpha} G_M(x,t|x_0)\right]\, ,
\label{GM.1}
\end{equation}
but now this equation is valid only in the range $x\le M$ with the boundary conditions
\begin{equation}
G_M(x=M,t|x_0)=0\, , \quad {\rm and}\quad G_M(x\to -\infty,t|x_0)\to 0\, ,
\label{GM_bc.1}
\end{equation}
and the initial condition $G_M(x,t=0|x_0)= \delta(x-x_0)$. Previously, in the free space, we found the simple scaling solution 
in Eqs. (\ref{eq:scaling_ansatz_propagator}) and (\ref{eq:propagator_free}). But this simple scaling approach does not work for the
solution in the constrained space $x\le M$ that satisfies the absorbing boundary condition $G_M(x=M,t|x_0)=0$. One needs
to solve the full boundary and the initial value problem which is considerably harder. However, we show below
that since we are finally interested in the solution $G_M(x=M-\epsilon, t|x_0)$ for $x=M-\epsilon$ close to the
absorbing boundary at $M$, one can bypass the problem of solving the full partial differential equation \eqref {GM.1} and
instead compute $G_M(x=M-\epsilon, t|x_0)$ by relating it to the survival probability discussed in the previous subsection.

To establish this relation, we first rewrite Eq. (\ref{GM.1}) as a continuity equation
\begin{equation}
\label{cont.1}
\frac{\partial G_M(x,t|x_0)}{\partial t}= - \frac{\partial J(x,t|x_0)}{\partial x}\, , \quad {\rm for}\quad x\le M
\end{equation}
where the current density $J(x,t|x_0)$ is given by (see Eq. (\ref{fp.1}))
\begin{equation} 
\label{eq:current_expression}
J(x,t|x_0)= -D(x)\, \frac{\partial G_M(x,t|x_0)}{\partial x}- U'(x)\, G_M(x,t|x_0)
\end{equation}
with the space dependent diffusion constant $D(x)= |x|^{-\alpha}$ and the drift potential $U(x)=|x|^{-\alpha}$ given
in Eq. (\ref{du.1}). Now, by definition, integrating $G_M(x,t|x_0)$ over $x\in [-\infty,M]$ precisely gives the survival probability
$Q(x_0,t|M)$ derived in the previous subsection  
\begin{equation}
Q(x_0,t|M)= \int_{-\infty}^M G_M(x,t|x_0)\, dx\, .
\label{GM_surv.1}
\end{equation}
Taking a derivative with respect to $t$ and using the continuity equation \eqref{cont.1} gives
\begin{equation}
\frac{d Q(x_0,t|M)}{dt}= - J(M,t|x_0)\, ,
\label{Q_current.1}
\end{equation}
where we used the fact that the current density vanishes as $x\to -\infty$, i.e., $J(x\to -\infty, t|x_0)=0$. 
Now close to $x=M$, due to the absorbing boundary condition $G_M(x=M,t|x_0)=0$, we can neglect
the second term in Eq. (\ref{eq:current_expression}) and approximate the current density by
\begin{equation}
J(x,t|x_0)\approx - D(M)\, \frac{\partial G_M(x,t|x_0)}{\partial x} \, .  
\label{Jx_approx.1}
\end{equation}
Using this in Eq. (\ref{Q_current.1}) we get, for $x$ close to $M$,
\begin{equation}
\frac{\partial G_M(x,t|x_0)}{\partial x}\approx \frac{1}{D(M)}\, \frac{d Q(x_0,t|M)}{dt}\, .
\label{GM_der.1}
\end{equation}
Integrating further locally over $x\in [M-\epsilon,M]$ and using $G_M(x=M,t|x_0)=0$ establishes the relation 
\begin{equation}
G_M(M-\epsilon,t|x_0)\approx -\frac{\epsilon}{D(M)}\, \frac{d Q(x_0,t|M)}{dt}= \frac{\epsilon}{D(M)}\, F(x_0,t|M) ,
\label{relation.1}
\end{equation}
where $F(x_0,t|M)= -dQ(x_0,t|M)/dt$ is the first-passage probability density to the level $M$, starting from $x_0$.
It is convenient also to compute the Laplace transform of $G_M(M-\epsilon,t|x_0)$ with respect to $t$ defined as
\begin{equation}
\tilde G_M(M-\epsilon,s|x_0)= \int_0^{\infty} dt\, e^{-s\, t}\, G_M(M-\epsilon,t|x_0)\, .
\label{GM_laplace.1}
\end{equation}
Taking Laplace transform of \eqref{relation.1} with respect to $t$, integrating $e^{-s\,t}\, dQ(x_0,t|M)/dt$ by parts and using
$Q(x_0,t=0|M)=1$, we get 
\begin{equation}
\tilde G_M(M-\epsilon,s|x_0)\approx \frac{\epsilon}{D(M)}\, \left[1- s\, \tilde Q(x_0,s|M)\right]\, .
\label{GM_laplace.2}
\end{equation}
Finally, setting $x_0=0$ as needed for the left part of \eqref{eq:joint_distribution_M_t_M} and using the result in Eq. (\ref{Q0.1})
we get, to leading order in $\epsilon$
\begin{equation}
\label{eq:first_passage_laplace}
\tilde G_M(M-\epsilon,s|0)= \frac{\epsilon\, \Gamma(\mu)}{2\, \mu^{\mu}\, D(M)\, \sqrt{M}}\,
\frac{s^{-\frac{\mu}{2}}}{r_\mu(s^\mu M)} + O(\epsilon^2)\, .
\end{equation}
Inverting this Laplace transform formally, we then get, to leading order in $\epsilon$,  the left part of the probability weight in 
Eq. (\ref{eq:joint_distribution_M_t_M})
\begin{equation}
G_M(M-\epsilon, t_M|0)= \frac{\epsilon\, \Gamma(\mu)}{2\, \mu^{\mu}\, D(M)\, \sqrt{M}}\, {\cal L}^{-1}_{s\to t_M}
\left[\frac{s^{-\frac{\mu}{2}}}{r_\mu(s^\mu M)}\right] + O(\epsilon^2) \, ,
\label{linv_left}
\end{equation}  
where $r_\mu(z)$ is defined in \eqref{eq:r_function}.

\subsubsection{The joint distribution $P(M,t_M|T)$}

Having computed both the left and the right part of the probability weight in Eq. (\ref{eq:joint_distribution_M_t_M}) to leading order
in $\epsilon$, we then take their product to get (absorbing the unimportant constants in the proportionality factor)
\begin{equation}
\label{product.1}
P(M,t_M|T) \propto \frac{\epsilon^2}{D(M)}\, \frac{1}{M}\, {\cal L}^{-1}_{s\to t_M} \left[\frac{s^{-\frac{\mu}{2}}}{r_\mu(s^\mu M)}\right]\,
{\cal L}^{-1}_{s\to (T-t_M)}\left[ \frac{\p_M\left[\sqrt{M}r_\mu\left(s^\mu M\right)\right]}
{s\, r_\mu\left(s^\mu M\right)}\right]\, ,
\end{equation}
where we recall that $r_\mu(z)$ is given in Eq. (\ref{eq:r_function}). Now comes the important physical observation.
The l.h.s of \eqref{product.1} represents the joint probability density, and hence $P(M,t_M|T) dM\, dt_M$
represents the probability weight of all paths satisfying the necessary constraints. On the r.h.s
the term ${\cal L}^{-1}_{s\to t_M} \left[\frac{s^{-\frac{\mu}{2}}}{r_\mu(s^\mu M)}\right]$ 
represents the probability density to arrive at $M-\epsilon$. So, we need to multiply
by the volume factor $dM$ to make it a probability. However, there is no $dt_M$ term on the r.h.s 
of \eqref{product.1}.
In fact this $dt_M$ indeed emerges naturally by identifying
$\epsilon^2/D(M)$ as proportional to the time interval $dt_M$ since $D(M)$, by definition,
represents the local diffusion constant near $x=M$. Hence, we arrive at the result
\begin{equation}
P(M,t_M|T)\, dM\, dt_M = \frac{C}{M}\, {\cal L}^{-1}_{s\to t_M} \left[\frac{s^{-\frac{\mu}{2}}}{r_\mu(s^\mu M)}\right]\,
{\cal L}^{-1}_{s\to (T-t_M)}\left[ \frac{\p_M\left[\sqrt{M}\, r_\mu\left(s^\mu M\right)\right]}
{s\, r_\mu\left(s^\mu M\right)} \right]\, dM\, dt_M \, ,
\label{joint_dist.1}
\end{equation}
where $C$ is an overall constant to be fixed from the normalization condition:
$\int_0^{\infty} dM \int_0^T dt_M\, P(M,t_M|T)=1$.

The convolution structure in Eq. (\ref{joint_dist.1}) suggests naturally to take its double Laplace transform
\begin{equation}
\tilde P(M,p|s)= \int_0^{\infty} dT\, e^{-s\, T}\, \int_0^T dt_M\, e^{-p\, t_M}\, P(M,t_M|T)\, .
\label{dlt.1}
\end{equation}
Taking this double Laplace transform in Eq. (\ref{joint_dist.1}) we immediately get
\begin{equation}
\tilde P(M,p|s)=\int_0^{\infty} dT\, e^{-s\, T}\, \int_0^T dt_M\, e^{-p\, t_M}\, P(M,t_M|T)=
 \frac{C}{M}\, \frac{(s+p)^{-\mu/2}}{r_\mu\left((s+p)^\mu\, M\right)}\, 
\frac{\partial_M\left(\sqrt{M}\, r_\mu(s^\mu\,M)\right)}{s\, r_\mu(s^\mu\, M)}\, .
\label{joint_dist.2}
\end{equation}

\subsubsection{The marginal distribution $p_M(t_M|T)$ and the scaling function $f_M(\xi)$}

Having obtained the Laplace transform of the joint distribution in \eqref{joint_dist.2},
we next integrate it over $M$ to obtain the marginal distribution $p_M(t_M|T)$. This gives
\begin{equation}
\int_0^{\infty} dT\, e^{-s\, T}\, \int_0^T dt_M\, e^{-p\, t_M}\, p_M(t_M|T)=\frac{C\, (s+p)^{-\mu/2}}{s} \int_0^{\infty} \frac{dM}{M}\,
\frac{\partial_M\left(\sqrt{M}\, r_\mu(s^\mu\,M)\right)}{r_\mu\left((s+p)^\mu\, M\right)\, r_\mu(s^\mu\, M)}\, .
\label{marginal_tm.1}
\end{equation}
On the l.h.s, substituting the scaling form $p_M(t_M|T)= (1/T)\, f_M(t_M/T)$, we get
\begin{equation}
\int_0^{\infty} dT\, e^{-s\, T}\, \int_0^T dt_M\, e^{-p\, t_M}\, P_M(t_M|T)= \int_0^1 \frac{d\xi\, f_M(\xi)}{s+ p\, \xi}\, .
\label{marginal_lhs.1}
\end{equation}
On the r.h.s of Eq. (\ref{marginal_tm.1}) we make the change of variable $s^{\mu}\, M=y$. This gives, after absorbing
unimportant constants into $C$ and denoting it by $C_1$,
\begin{equation}
\int_0^1 \frac{d\xi\, f_M(\xi)}{s+ p\, \xi}= \frac{C_1}{s}\,\left(1+\frac{s}{p}\right)^{-\mu/2}\,
\int_0^{\infty} \frac{dy}{y} \frac{\partial_y\left(\sqrt{y}\, r_\mu(y)\right)}{r_\mu(y)\, r_\mu\left(\left(1+\frac{s}{p}\right)^{\mu}\, 
y\right)}\, .
\label{marginal_tm.2}
\end{equation}
We then set $s=-pz$ on both sides and obtain the Stieltjes transform
\begin{equation}
\int_0^1 \frac{d\xi\, f_M(\xi)}{z-\, \xi}= \frac{C_1}{z}\, \left(1-\frac{1}{z}\right)^{-\mu/2}\,
\int_0^{\infty} \frac{dy}{y} \frac{\partial_y\left(\sqrt{y}\, r_\mu(y)\right)}{r_\mu(y)\, r_\mu\left(\left(1-
\frac{1}{z}\right)^{\mu}\, y\right)}\,  .
\label{marginal_tm.3}
\end{equation}
To simplify the right hand side further, we make the change of variable $2\mu y^{1/{2\mu}}=x$. 
Using the definitions of $r_\mu(z)$ in Eq. (\ref{eq:r_function}) and $g_\mu(z)$ in Eq. (\ref{eq:auxiliary_function_g}), we get
after simple algebra
\begin{equation}
\label{marginal_tm_final}
    \int_0^1 \dd \xi\, \frac{f_M(\xi)}{z-\xi} = \frac{\mc{N}}{z}\int_0^{\infty} dx\, \frac{g_\mu'(x)}{g_\mu(x)}\,
\frac{1}{\left[g_\mu\left(\sqrt{1-\frac{1}{z}}\, x\right)\right]}\, ,
\end{equation}
where all the constants from the change of variables have been absorbed in the new constant $\mc{N}$.
We fix this constant from the normalization condition $\int_0^{1} f_M(\xi)\, d\xi=1$. Indeed, comparing the $1/z$
term in the large $z$ expansion on both sides of Eq. (\ref{marginal_tm_final}) and using this normalization condition we get
\begin{equation}
 \mc{N}\, \int_0^{\infty} dx\, \frac{g_\mu'(x)}{g_\mu^2(x)}= 1\, .
\label{norm.1}
\end{equation}
Carrying out the integral on the l.h.s trivially, we fix the normalization constant
\begin{equation}
\mc{N}= \lim_{x\to 0} g_\mu(x) = \frac{2^{\mu}}{\Gamma(1-\mu)}\, ,
\label{norm.2}
\end{equation}
where we used the definition of $g_\mu(x)$ and the small $x$ asymptotics of the Besssel function in 
Eq. (\ref{eq:bessel_second_small_argument}). The result in \eqref{marginal_tm_final} along with $\mc{N}$
in \eqref{norm.2} thus completes our derivation of the exact Stieltjes transform of $f_M(\xi)$ announced in
Eq. (\ref{eq:stieltjes_time_of_max}).

The Stieltjes transform in \eqref{marginal_tm_final}
can be inverted using the Sokhotski–Plemelj formula \eqref{IST.1} leading to the final result
\begin{equation}
    f_{M}(\xi) = \frac{\mc{N}}{\xi\, \pi}\, \Im\int_0^{\infty} \frac{\dd x}
{\left[g_\mu\left(-\ii\sqrt{\frac{1}{\xi}-1}\, x\right)\right]}\, \frac{ g'_\mu(x)}{g_\mu(x)}\, ,
\label{eq:f_M.1}
\end{equation}
that was announced before. We recall that $g_\mu(x)= x^\mu\left(I_\mu(x)+I_{-\mu}(x)\right)$.
Using the asymptotic behaviors of $g_\mu(x)$ in Eq. (\ref{gx_asymp}),
one can extract the asymptotic behaviors of the scaling function $f_M(\xi)$ at the two edges $\xi\to 0$ and $\xi\to 1$,
as announced in Eq. (\ref{fM_asymp}) (the details are provided in Appendix B). 

The formula \eqref{eq:f_M.1} is exact and valid for all $\xi\in(0,1)$. To plot it, we need to
perform the integral in  \eqref{eq:f_M.1} numerically. 
This, however, is rather hard because the modified Bessel functions of imaginary argument 
are related to standard Bessel functions that are highly oscillatory. 
The trick to allow a simple numerical evaluation of the integral above makes use of complex analysis. 
Indeed, since the integrand is an analytic function of $x$, we can deform the contour into the complex plane 
using Cauchy's theorem. 
We integrate the complexified integrand along a closed contour going from $0$ to $R>0$ along the real axis, up into the upper 
half-plane along the segment of the circle parametrized by $R\, e^{\ii \omega}$ and then back to the origin along a straight 
line. This contour encircles a wedge shaped domain $\gamma$ of angle $\omega$ as shown in Fig. \ref{fig:complex_contour}.
The integral along the closed path is $0$ by Cauchy's theorem because, as can be verified numerically, 
the function $g_\mu(x)$ does not have zeros in the right half plane $\Re x \geq 0$. 
Since the integrand also decays fast as $R\to +\infty$, the contribution along 
the segment of a circle vanishes and we obtain
\begin{align}\label{eq:integral_well_behaved}
    \int_0^{+\infty}\frac{\dd x}
{\left[g_\mu\left(-\ii\sqrt{\frac{1}{\xi}-1}\, x\right)\right]}\, 
\frac{\p_x g_\mu(x)}{g_\mu(x)} &= 
\lim_{R\to \infty} \int_0^{R\, e^{i\omega}} \frac{\dd z}   
{\left[g_\mu\left(-\ii\sqrt{\frac{1}{\xi}-1}\, z\right)\right]}\,
\frac{\partial_z g_\mu(z)}{g_\mu(z)} \nonumber \\
    & = \int_0^{ +\infty} \frac{\dd x \,e^{\ii \omega}}{\left[g_\mu\left(-\ii\,e^{\ii \omega}\sqrt{\frac{1}{\xi}-1}\, 
x\right)\right]}\, \frac{\p_x g_\mu(e^{\ii \omega}\,x)}{g_\mu(e^{\ii \omega}\, x)}\quad .
\end{align}
The plots in Fig. \ref{fig:f_M} have been obtained with the choice $\omega \approx \pi/4$ and one can see that the results are 
extremely good. This is because now the integrand in the last line in \eqref{eq:integral_well_behaved}
is exponentially convergent. 

\begin{figure}
    \centering
    \includegraphics[width=0.6\linewidth]{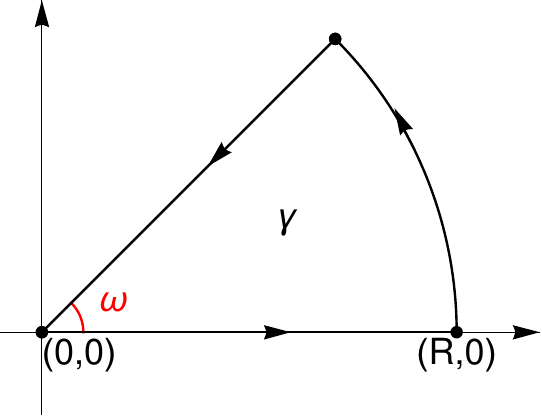}
    \caption{Wedge-shaped integration path used to numerically evaluate the integral \eqref{eq:f_M}. 
The integral around the domain $\gamma$ is zero because the integrand is analytic. 
In the limit $R\to +\infty$ the contribution from the circular part vanishes and one obtains the well behaved 
integral in \eqref{eq:integral_well_behaved}.}
    \label{fig:complex_contour}
\end{figure}

\section{Conclusions}\label{section:4}

In summary, we have studied a simple one dimensional model of a sluggish random walker with subdiffusive growth. 
In the continuum limit, this
model corresponds to a particle diffusing on a line with a space dependent diffusion constant $D(x)=|x|^{-\alpha}$
and a drift potential $U(x)=|x|^{-\alpha}$, where $\alpha>0$ parametrizes the dynamics. The case $\alpha=0$ corresponds
to normal diffusion. For $\alpha>0$, the dynamics becomes `sluggish' the further the particle 
goes away from the origin, leading
to a slow subdiffusive growth of the typical displacement $x\sim t^{\mu}$ for large $t$ where 
$0\le \mu=1/(\alpha+2)\le 1/2$.
In this paper, we have computed exactly the probability distributions of three main observables in the sluggish random walk model
of total duration $T$:
(i) the occupation time $t_+$ denoting the time spent on the positive side of the origin, (ii) the last passage time $t_{\rm l}$
through the origin before $T$, and (iii) the time $t_M$ at which the walker is maximally displaced on the positive side
of the origin. We showed that all three distributions admit the scaling forms
\begin{equation}
{\rm Prob.}[t_+=t|T]= \frac{1}{T}\, f_+\left(\frac{t}{T}\right)\, , \quad\quad
{\rm Prob.}[t_{\rm l}=t|T]= \frac{1}{T}\, f_{\rm l}\left(\frac{t}{T}\right)\, , \quad\quad
{\rm Prob.}[t_M=t|T]= \frac{1}{T}\, f_M\left(\frac{t}{T}\right)\, , \quad\quad\, .
\label{scaling_forms}
\end{equation}
In this paper, we have computed the three scaling functions
$f_+(\xi)$, $f_{\rm l}(\xi)$ and $f_M(\xi)$ analytically for all $\alpha\ge 0$ and have shown that they
all have convex shapes over the support $\xi\in [0,1]$.
For $\alpha=0$ (normal diffusion), they
are all identical and reduce to the celebrated arcsine form of L\'evy. However, we have shown that for any $\alpha>0$,
the three scaling functions are different from each other and depend nontrivially on $\alpha$. For $\alpha\ge 0$, i.e.,
for $0<\mu\le 1/2$, while all three of them
diverge as $\sim \xi^{\mu-1}$ as $\xi\to 0$ (albeit with different $\alpha$ dependent prefactors), 
they diverge differently as $\xi$ approaches the upper edge $\xi=1$,
\begin{equation}
f_+(\xi) \sim (1-\xi)^{\mu-1}\, , \quad\quad f_{\rm l}(\xi) \sim (1-\xi)^{-\mu}\, , \quad\quad f_M(\xi)\sim (1-\xi)^{-1/2}\, .
\label{diverge_xi1}
\end{equation}
We also performed numerical simulations and found excellent agreement with our analytical predictions. 
Our result provides a rare example of a non-Brownian process for which the distributions of all these
three observables can be computed exactly.

This work raises many interesting open questions. The occupation time 
$t_+= \int_0^T \Theta\left(x(t)\right)\, dt$ studied here
is a functional of the sluggish walker process $x(t)$ of duration $T$. One can also study a more general functional of the
form $Y= \int_0^T V\left(x(t)\right)\, dt$. In Appendix B, we have derived the generalized 
backward Feynman-Kac equation
to study the Laplace transform of the distribution of $Y$ for general $V(x)$ (see Eq. (\ref{bfk_A1})).
One can use this equation to study, e.g., the distribution of the area under the sluggish process up to $T$ by
choosing $V(x)=x$. For the diffusion process $(\alpha=0)$, the distribution of the area is a pure Gaussian
since $\int_0^T x(t)\, dt$ is a linear functional of $x(t)$ which itself is Gaussian. However, for $\alpha>0$,
the process $x(t)$ is non-Gaussian and consequently, the distribution of the area will also be non-Gaussian and 
nontrivial.
It would be interesting to compute the distribution of the area or other functionals explicitly for the
sluggish walker.

It would be interesting to compute the distributions of $t_+$, $t_{\rm l}$ and $t_M$ in higher 
dimensions. For instance, in $d=2$, one can define the occupation time $t_{\cal D}$ as the time spent in the domain $\cal D$ by 
the sluggish walker of duration $T$. Can one compute the distribution of $t_{\cal D}$ for some choice of domain
$\cal D$? For a Brownian 
motion ($\alpha=0$), some results are known for $t_{\cal D}$ for certain choices of $\cal D$, for example, when $\cal D$ 
represents a wedge of angle $\omega$~\cite{CD2003}. Can one extend these results to $\alpha>0$? 
Another interesting question concerns the 
study of constrained random walkers in one dimension. For example, for a Brownian motion of duration $T$, 
several variants of 
constrained Brownian motions such as bridges, excursions, meanders etc, have been well studied in the 
literature~\cite{MRKY2008,MC2005}. It would be 
interesting to study the properties of the sluggish version $(\alpha>0)$ of such constrained random walks. 
Finally, since
the three observables $t_+$, $t_{\rm l}$ and $t_M$ are of natural interest in the context of financial time series 
(e.g. the stock prices), it would
be interesting to see if our results for the sluggish walker model can be applied to real financial data.

\appendix

\section{Derivation of the backward Feynman-Kac equation}

In this Appendix we provide a derivation of the backward Feynman-Kac equation \eqref{eq:backward_equation} in the
main text. Actually, 
one can derive the equation for a more general functional of the form
\begin{equation}
Y= \int_0^T V\left(x(t)\right)\, \dd t\, ,
\label{funct.A1}
\end{equation}
where $x(t)$ represents the sluggish random walk trajectory and $V(x)$ is an arbitrary function. The occupation time $t_+$
corresponds to the special choice $V(x)=\Theta(x)$. Let $P(Y|T,x_0)$ denote the distribution of $Y$ given
that the trajectory starts at $x_0$ and is of duration $T$. We define its Laplace transform with 
respect to $Y$ 
\begin{equation}
\phi(x_0,T|p)= \braket{e^{-p\, \int_0^T V\left(x(t)\right)\, dt}}_{x_0} = \int_0^{\infty}\dd Y\, e^{-p\,Y}\,P(Y|T,x_0) \,.
\label{lt_A2}
\end{equation}

It is easier to start from the discrete time lattice version of the model, where we choose the lattice constant $a$
and time step $\Delta T$. In a time step $\Delta T$, the walker, starting from $x_0$, hops to
$x_0\pm a$ with probability $\left(\left(\frac{|x_0|}{a}\right)^{\alpha}+2\right)^{-1}$ and 
stays at $x_0$ with the complementary probability
$1- 2\, \left(\left(\frac{|x_0|}{a}\right)^{\alpha}+2\right)^{-1}$. 
We next consider the quantity $Q_p(x,T+\Delta T, x_0)$ and split
the time interval $[0, T+\Delta T]$ into two parts: (i) $[0,\Delta T]$ (the first step) and (ii) $[\Delta T, T+\Delta T]$.
 Considering the stochastic moves during the first step $\Delta T$, we can then write the following
recursion relation
\begin{equation}
\phi(x_0,T+\Delta T|p)= e^{-p\, V(x_0)\, \Delta T}\, \left[ 
\frac{1}{\left(\frac{|x_0|}{a}\right)^{\alpha}+2}\, \left(\phi(x_0-a,T|p)
 +\phi(x_0+a,T|p)\right) + \left(1- \frac{2}{\left(\frac{|x_0|}{a}\right)^{\alpha}+2}\right)\, \phi(x_0,T|p)\right]\, .
\label{recur_A1}
\end{equation}
This is easy to understand. The contribution to $Y$ in the first step is simply $V(x_0)\,\Delta T$ and after this first step
one again follows the trajectory starting at the new positions at step $\Delta T$. 
We then take the continuum limit $a\to 0$, $\Delta T\to 0$ with the ratio $a^{\alpha+2}/\Delta T$ fixed
to unity. Then one immediately arrives at the backward Feynman-Kac equation
\begin{equation}
 \p_T \phi(x_0,T|p)= \frac{1}{|x_0|^{\alpha}}\, \p_{x_0}^2 \phi(x_0,T|p)- p\, V(x_0)\, \phi(x_0,T|p)\, .
\label{bfk_A1}
\end{equation}
Choosing $V(x)=\Theta(x)$ thus gives \eqref{eq:backward_equation} in the main text.

\section{Derivation of the edge behaviors of $f_M(\xi)$ as $\xi\to 0$ and $\xi\to 1$}

In this Appendix, we work out the asymptotic behaviors of the scaling function $f_M(\xi)$ in Eq. (\ref{fM_asymp})
near the two edges $\xi\to 0$ and $\xi\to 1$. 
While one can analyse these behaviors starting from the formal
explicit expression in \eqref{eq:f_M}, they turn out to be easier to extract from the Stieltjes transform
in Eq. (\ref{eq:stieltjes_time_of_max}) as we show below.

\vskip 0.3cm

\noindent {\bf The behavior of $f_M(\xi)$ near $\xi\to 1$.}
We start with the edge behavior $\xi\to 1$, as it is somewhat simpler. To extract this limit 
we need to investigate the $z\to 1$ limit of the Stieltjest transform because the l.h.s of 
Eq. (\ref{eq:stieltjes_time_of_max}) near $z=1$ is dominated
by the $\xi\to 1$ behavior of $f_M(\xi)$. We set $z=1+\delta$ on the l.h.s of \eqref{eq:stieltjes_time_of_max}
and make the change of variable $1-\xi= u \delta$. Then in the limit $\delta\to 0$, Eq. (\ref{eq:stieltjes_time_of_max})
reduces to
\begin{equation}
\int_0^{1/\delta} \frac{f_M(1-u \delta)\, du}{1+u}\approx \mc{N}\, \int_0^1 \frac{dx}{g_\mu\left(\sqrt{\delta}\, x\right)}\,
\frac{g_\mu'(x)}{g_\mu(x)}\, .
\label{xi1_A2}
\end{equation}
Next on the r.h.s of \eqref{xi1_A2}, we rescale $\sqrt{\delta} x=y$ and take the $\delta\to 0$ limit.
Using the asymptotic large $x$ behavior of $g_\mu(x)\approx \sqrt{2/\pi}\, x^{\mu-1}\, e^{x} $ (see Eq. (\ref{gx_asymp})) in
the integrand, we get the leading behavior of the r.h.s
\begin{equation}
\int_0^{1/\delta} \frac{f_M(1-u \delta)\, du}{1+u}\approx \frac{\mc{N}}{\sqrt{\delta}}\, \int_0^{\infty} \frac{dy}{g_\mu(y)\, .
\label{xi1_A2.1}}
\end{equation}
This indicates that as $\delta\to 0$, we must have $f_M(1-u\, \delta)\to B_\mu/\sqrt{u\,\delta}$ in order that both sides
of Eq. (\ref{xi1_A2.1}) have the same leading power of $\delta$ as $\delta\to 0$, i.e., $\delta^{-1/2}$. 
Matching
the prefactor on both sides we get
\begin{equation}
B_\mu\, \int_0^{\infty} \frac{du}{\sqrt{u}\, (1+u)} = \mc{N}\, \int_0^{\infty} \frac{\dd y}{g_\mu(y)}\, . 
\label{xi1_A2.2}
\end{equation}
Performing the integral on the l.h.s and using $\mc{N}= 2^{\mu}/\Gamma(1-\mu)$ from \eqref{eq:normalisation_M} then gives
\begin{equation}
B_\mu= \frac{2^\mu}{\Gamma(1-\mu)\,\pi}\, \int_0^{\infty}\frac{\dd y}{g_\mu(y)}\, ,
\label{Bmu_deriv.1}
\end{equation} 
as announced in Eq. (\ref{Bmu.1}).

\vskip 0.3cm

\noindent {\bf The behavior of $f_M(\xi)$ near $\xi\to 0$.} We now turn to the other edge $\xi\to 0$. In order to
extract the behavior of $f_M(\xi)$ in this regime, we need to analyse the Stieltjes transform \eqref{eq:stieltjes_time_of_max}
in the limit $z\to 0$. We now set $z=-\delta$ with $\delta\to 0$ on both sides of Eq. (\ref{eq:stieltjes_time_of_max}).
On the l.h.s. , we rescale $\xi= u\,\delta$. This gives
\begin{equation}
\int_0^{1/\delta} \frac{f_M(u \delta)\, du}{1+u}\approx
\frac{\mc{N}}{\delta}\, \int_0^{\infty} \frac{dx}{g_\mu\left(\frac{x}{\sqrt{\delta}}\right)}\, \frac{g_\mu'(x)}{g_\mu(x)}=
\frac{\mc{N}}{\sqrt{\delta}}\, \int_0^{\infty} \frac{dy}{g_\mu(y)}\, \frac{g_\mu'\left(\sqrt{\delta}\, y\right)}{g_\mu\left(
\sqrt{\delta}\, y\right)} \,
\label{xi0.A2}
\end{equation}
where in arriving at the last integral we made a change of variable $x=  \sqrt{\delta}\, y$.
Now, we take the $\delta\to 0$ limit on the r.h.s of \eqref{xi0.A2}. For this, we will use the small
argument asymptotic behavior of $g_\mu(x)$ in \eqref{gx_asymp}. Indeed, for small $x$, we have the
leading asymptotic behavior
\begin{equation}
\frac{g_\mu'(x)}{g_\mu(x)}\approx \frac{2\mu\, 2^{-2\mu}\, \Gamma(1-\mu)}{\Gamma(1+\mu)}\, x^{2\mu-1}\, .
\label{xi0.A2.1}
\end{equation}
Substituting this behavior inside the integrand on the r.h,s of \eqref{xi0.A2} we get
\begin{equation}
\int_0^{1/\delta} \frac{f_M(u \delta)\, du}{1+u}\approx {\mc{N}}\, 
\frac{2\mu\, 2^{-2\mu}\, \Gamma(1-\mu)}{\Gamma(1+\mu)}\, \delta^{\mu-1}\, 
\int_0^{\infty} \frac{dy\, y^{2\mu-1}}{g_\mu(y)}\, .
\label{xi0.A2.2}
\end{equation}
Once again, in order that the leading power of $\delta$ matches on both sides of \eqref{xi0.A2.2}, we must have
$f_M(u\, \delta) \to A_\mu (u\, \delta)^{\mu-1}$ as $\delta\to 0$. Matching the prefactors we get
\begin{equation}
A_\mu \, \int_0^{\infty} \frac{u^{\mu-1}\, du}{u+1}= \frac{\mc{N}\, 2\mu\, 2^{-2\mu}\, \Gamma(1-\mu)}{\Gamma(1+\mu)}\, 
\int_0^{\infty} \frac{dy\, y^{2\mu-1}}{g_\mu(y)}\, .
\label{xi0.A2.3}
\end{equation}
Using $\mc{N}= 2^\mu/\Gamma(1-\mu)$ and simplifying we get
\begin{equation}
A_\mu= \frac{2\mu\, 2^{-\mu}}{\Gamma(\mu)\,\Gamma(1-\mu)\,\Gamma(1+\mu)}\,
\int_0^{\infty} \frac{dy\, y^{2\mu-1}}{g_\mu(y)}\, ,
\label{Amu_deriv.1}
\end{equation}
as announced in Eq. (\ref{Amu.1}).

\section*{Acknowledgements}

We acknowledge support from ANR Grant No. ANR-23-CE30-0020-01 EDIPS.


\begin{thebibliography}{99}


\bibitem{Feller1950} W. Feller, {\em Introduction to Probability Theory and Its Applications}, John Wiley Sons, New 
York (1950).

\bibitem{Levy1940} P. L\'evy, {\em Sur certains processus stochastiques homog\'enes}, Compos. Math. {\bf 7}, 283 (1940).


\bibitem{dale1980}
C. Dale, R. Workman,
\textit{The Arc Sine Law and the Treasury Bill Futures Market}, 
Financial Analysts Journal {\bf 36}, 71 (1980) (http://www.jstor.org/stable/4478403).

\bibitem{Majumdar2005} S. N. Majumdar, {\em Brownian Functionals in Physics and Computer Science}, Curr. Sci. {\bf 
89}, 2076 (2005).

\bibitem{dey2022}
R. Dey, A. Kundu, B. Das, A. Banerjee, 
\textit{Experimental verification of arcsine laws in mesoscopic nonequilibrium systems},
Phys. Rev. E,  \textbf{106}, 054113 (2022).











\bibitem{Redner2001} S. Redner, {\em A guide to first-passage processes}, Cambridge University Press, Cambridge (2001).

\bibitem{Bray2013} A. J. Bray, S. N. Majumdar, G. Schehr, {\em Persistence and first-passage properties in 
nonequilibrium systems}, Adv. in Phys. {\bf 62}, 225 (2013).

\bibitem{Majumdar2010a} S. N. Majumdar, {\em Universal First-passage Properties of Discrete-time Random Walks and 
Levy Flights on a Line: Statistics of the Global Maximum and Records}, Physica A {\bf 389}, 4299 (2010).




\bibitem{Andersen1954} E. Sparre Andersen, {\em On the fluctuations of sums of random variables}, Math. Scand. {\bf 
1}, 263 (1954).


\bibitem{Watanabe1995} S. Watanabe, {\em Generalized arc-sine laws for one-dimensional diffusion processes and random walks}, Proc. Symp. Pure Math. {\bf 57}, 157 (1995).


\bibitem{Lamperti1958} J. Lamperti,
\textit{An occupation time theorem for a class of stochastic processes},
Trans. Am. Math. Soc. \textbf{88}, 380 (1958).

\bibitem{Godreche2001} C. Godrèche, J. M. Luck, {\em Statistics of the occupation time of renewal processes}, J. 
Stat. Phys. {\bf 104}, 489 (2001).

\bibitem{Burov2011} S. Burov, E. Barkai, {\em Residence Time 
Statistics for $N$ Renewal Processes},
Phys. Rev. Lett. {\bf 107}, 170601 (2011).

\bibitem{Dhar1999} A. Dhar, S. N. Majumdar, {\em Residence time distribution for a class of Gaussian Markov 
processes}, Phys. Rev. E {\bf 59}, 6413 (1999).

\bibitem{DeSmedt2001} G. De Smedt, C. Godrèche, J. M. Luck, {\em Statistics of the occupation time for a class 
of Gaussian Markov processes}, J. Phys. A: Math. Gen. {\bf 34}, 1247 (2001).

\bibitem{Dornic1998} I. Dornic, C. Godr\`eche, {\em Large deviations and nontrivial exponents in
coarsening systems}, J. Phys. A: Math. Gen. {\bf 31}, 5413 (1998

\bibitem{Newman1998} T. J. Newman, Z. Toroczkai, {\em Diffusive persistence and the sign-time distribution}, Phys. Rev. E {\bf 58}, R2685 (1998).

\bibitem{Drouffe1998} J.-M. Drouffe, C. Godrèche, {\em Stationary definition of persistence for finite 
temperature phase ordering}, J. Phys. A {\bf 31}, 9801 (1998).


\bibitem{Toroczkai1998} Z. Toroczkai, T. J. Newman, S. Das Sarma, {\em Sign-time distributions for interface 
growth}, Phys. Rev. E {\bf 60}, R1115 (1998).


\bibitem{Baldassarri1999} A. Baldassarri, J.-P. Bouchaud, I. Dornic, C. Godrèche, {\em Statistics of persistent 
events: An exactly soluble model}, Phys. Rev. E {\bf 59}, R20 (1999).


\bibitem{Brokmann2003} X. Brokmann, J.-P. Hermier, G. Messin, P. Desbiolles, J.-P. Bouchaud, M. Dahan, {\em 
Statistical Aging and Nonergodicity in the Fluorescence of Single Nanocrystals}, Phys. Rev. Lett. {\bf 90}, 120601 
(2003).


\bibitem{Margolin2005} G. Margolin, E. Barkai, {\em Nonergodicity of Blinking
Nanocrystals and Other L\'evy-Walk Processes}, Phys. Rev. Lett.
{\bf 94}, 080601 (2005).


\bibitem{Majumdar2002a} S. N. Majumdar, A. Comtet, {\em Local and the Occupation Time of a Particle Diffusing in a 
Random Medium}, Phys. Rev. Lett. {\bf 89}, 060601 (2002).

\bibitem{Sabhapandit2006} S. Sabhapandit, S.N. Majumdar, A. Comtet, “ Statistical
Properties of Functionals of the Paths of a Particle Diffusing in a One-Dimensional Random
Potential”, Phys. Rev. E v-73, 051102 (2006)

\bibitem{MD2002} S. N. Majumdar, D. S. Dean, {\em Exact occupation time distribution
in a non-Markovian sequence and its relation to spin
glass models}, Phys. Rev. E {\bf 66}, 041102 (2002).

\bibitem{Bel2005} G. Bel, E. Barkai, {\em Weak ergodicity breaking in the
continuous-time random walk}, Phys. Rev. Lett. {\bf 94}, 240602
(2005).

\bibitem{Barkai2006} E. Barkai, {\em Residence Time Statistics for Normal and Fractional
Diffusion in a Force Field}, J. Stat. Phys. {\bf 123}, 883 (2006).

\bibitem{RA2016} H. J. O. Boutcheng, T. B. Bouetou, T. W. Burkhardt, A. Rosso,
A. Zoia, K. T. Crepin, {\em Occupation time statistics of the random
acceleration model}, J. Stat. Mech. 053213 (2016).

\bibitem{Sadhu2018} T. Sadhu, M. Delorme, K. J. Wiese, {\em Generalized arcsine
laws for fractional Brownian motion}, Phys. Rev. Lett. {\bf 120},
040603 (2018).

\bibitem{denHollander2019} F. den Hollander, S. N. Majumdar, J. M. Meylahn, H.
Touchette, {\em Properties of additive functionals of Brownian motion
with resetting}, J. Phys. A: Math. Theor. {\bf 52}, 175001 (2019).

\bibitem{SK2019}
P. Singh, A. Kundu, {\em Generalised ‘Arcsine’ laws for run-and-tumble particle in one
dimension}, J. Stat. Mech. 083205 (2019).

\bibitem{MS2023} S. Mukherjee, N. R. Smith, {\em Dynamical phase transition in the occupation fraction
statistics for noncrossing Brownian particle}, Phys. Rev. E {\bf 107}, 064133 (2023).

\bibitem{MLS2024}S. Mukherjee, P. Le Doussal, N. R. Smith, {\em Large deviations in statistics of the
local time and occupation time for a run and tumble particle}, Phys. Rev. E {\bf 110},
024107 (2024).



\bibitem{Bressloff20}
P. C. Bressloff, {\em Occupation time of a run-and-tumble particle
with resetting}, Phys. Rev. E {\bf 102}, 042135 (2020).

\bibitem{Radice2020}
M. Radice, M. Onofri, R. Artuso, G. Pozzoli, {\em Statistics of
occupation times and connection to local properties of nonhomogeneous
random walks}, Phys. Rev. E {\bf 101}, 042103 (2020).



\bibitem{Kay2023}
T. Kay, L. Giuggioli, {\em
Extreme value statistics and Arcsine laws of Brownian motion in the presence of a
permeable barrier},
J. Phys. A: Math. Theor. {\bf 56}, 345002 (2023).


\bibitem{Nikeghbali2013} A. Nikeghbali, E. Platen, {\em A reading guide for last passage times with financial 
applications in view}, Finance Stochastics {\bf 17}, 615 (2013).

\bibitem{Leung2004} B. H. Leung, {\em A novel model on phase noise of ring oscillator based on last passage time}, 
IEEE Trans. Circuits Syst. I: Regular Papers, {\bf 51}, 471 (2004).


\bibitem{Hwang2006} C.-O. Hwang, J. A. Given, {\em Last-passage Monte Carlo algorithm for mutual capacitance}, Phys. 
Rev. E {\bf 74}, 027701 (2006).


\bibitem{yu2021}
U. Yu, Y.-M. Lee, C.-O. Hwang,
\textit{Last-passage Monte Carlo Algorithm for Charge Density on a Conducting Spherical Surface},
J. Sci. Comput. \textbf{88}, 82 (2021).

\bibitem{Barkar2009} C. T. Barkar, M. J. Newby, {\em Optimal non-periodic inspection for a multivariate degradation 
model}, Reliab. Eng. Syst. Saf. {\bf 94}, 33 (2009).

\bibitem{robson2014}
S. Robson, B. Leung, G. Gong,
\textit{Truly Random Number Generator Based on a Ring Oscillator Utilizing Last Passage Time},
IEEE Trans. Circuits Syst. II: Express Briefs \textbf{61}, 937 (2014).

\bibitem{clauset2015}
A. Clauset, M. Kogan, S. Redner,
\textit{Safe leads and lead changes in competitive team sports},
Phys. Rev. E \textbf{91}, 6 062815 (20015).


\bibitem{Comtet2020} A. Comtet, F. Cornu, G. Schehr, {\em Last-passage time for linear diffusions and application to 
the emptying time of a box}, J. Stat. Phys. {\bf 181}, 1565 (2020).




\bibitem{Majumdar2020} S. N. Majumdar, A. Pal, G. Schehr, {\em Extreme value statistics of correlated random variables: A pedagogical review}, Phys. Rep. {\bf 840}, 1 (2020).

\bibitem{Majumdar2024} S. N. Majumdar, G. Schehr, {\em Statistics of Extremes and Records in Random Sequences}, Oxford University Press (2024).




\bibitem{Shepp1979} L. A. Shepp, {\em The joint density of the maximum and its location for a Wiener process with drift}, J. Appl. Proba. {\bf 16}, 423 (1979).

\bibitem{Buffet2003} E. Buffet, {\em On the time of the maximum of Brownian motion with drift}, J. Appl. Math. Stoch. Anal. {\bf 16}, 201 (2003).

\bibitem{Baz2004} J. Baz, G. Chacko, {\em Financial derivatives: pricing, applications, and mathematics}
(Cambridge University Press, 2004).

\bibitem{Randon2007} J. Randon-Furling, S. N. Majumdar, {\em Distribution of the time at which the deviation
of a Brownian motion is maximum before its first-passage time}, J. Stat. Mech. P10008
(2007).

\bibitem{MB2008} S. N. Majumdar, J.-P. Bouchaud, {\em Optimal time to sell a stock in the Black–Scholes model},
Quant. Financ. {\bf 8}, 753 (2008).

\bibitem{CB2014} R. Chicheportiche, J.-P. Bouchaud. {\em Some applications of first-passage
ideas to finance} (book chapter in {\em First-Passage Phenomena and Their Applications} ed. by: R. Metzler, G. Oshanin and S. Redner, World Scientific) 447 (2014).


\bibitem{MRKY2008} S. N. Majumdar, J. Randon-Furling, M. J. Kearney, M. Yor, {\em On the time to reach
maximum for a variety of constrained Brownian motions}, J. Phys. A: Math. Theor. {\bf 41}, 365005
(2008).

\bibitem{Schehr2010} G. Schehr, P. Le Doussal, {\em Extreme value statistics from the real space renormalization 
group: Brownian motion, Bessel processes and continuous time random walks}, J. Stat. Mech. P01009 (2010).


\bibitem{RMC2009} J. Randon-Furling, S. N. Majumdar, A. Comtet, {\em
Convex Hull of N planar Brownian Motions: Application to Ecology},
Phys. Rev. Lett. {\bf 103}, 140602 (2009).

\bibitem{MCR2010} S. N. Majumdar, A. Comtet, J. Randon-Furling, {\em Random Convex Hulls and Extreme Value
Statistics}, J. Stat. Phys. {\bf 138}, 955 (2010).

\bibitem{CBM15.1} M. Chupeau, O. Benichou, S.N. Majumdar, {\em Mean perimeter of the convex hull of a 
random walk in a semi-infinite medium}, Phys. Rev. E, {\bf 92}, 022145 (2015).

\bibitem{CBM15.2} M. Chupeau, O. Benichou, S.N. Majumdar, {\em Convex hull of a Brownian motion in
confinement}, Phys. Rev. E, {\bf 91}, 050104 (2015).

\bibitem{RMR2011} A. Reymbaut, S. N. Majumdar, A. Rosso , {\em The convex hull for a random acceleration
process in two dimensions}, J. Phys. A: Math. Theor., {\bf 44}, 415001 (2011).

\bibitem{Dumonteil2013} E. Dumonteil, S.N. Majumdar, A. Rosso, A. Zoia, {\em Spatial extent of an outbreak in
animal epidemics},  Proc. Natl. Acad.
Sci. USA \text{\bf{110}}, 4239 (2013).

\bibitem{MMSS2021} S. N. Majumdar, F. Mori, H. Schawe, G. Schehr, {\em Mean perimeter and area of the
convex hull of a planar Brownian motion in the presence of resetting}, Phys. Rev. E {\bf
103}, 022135 (2021).

\bibitem{HMSS2020} A. K. Hartmann, S. N. Majumdar, H. Schawe, G. Schehr,
{\em The convex hull of the run-and-tumble particle in a plane},
J. Stat. Mech. 053401 (2020).


\bibitem{SKMS2022} P. Singh, A. Kundu, S. N. Majumdar, H. Schawe, {\em Mean area of the convex hull of a run
and tumble particle in two dimensions}, J. Phys. A.: Math. Theor. {\bf 55}, 225001
(2022).

\bibitem{MMS2019} F. Mori, S. N. Majumdar, G. Schehr, {\em Time between the maximum and the minimum of a
stochastic process}, Phys. Rev. Lett. {\bf 123}, 200201 (2019).

\bibitem{MMS2020} F. Mori, S. N. Majumdar, G. Schehr, {\em Distribution of the time between maximum and
minimum of random walks}, Phys. Rev. E, {\bf 101}, 052111 (2020).


\bibitem{MRZ2010} S. N. Majumdar, A. Rosso, A. Zoia, {\em Time at which the Maximum of a Random
Acceleration Process is reached}, J. Phys. A: Math. Theor. {\bf 43}, 115001 (2010).


\bibitem{MRZ2010a} S. N. Majumdar, A. Rosso, A. Zoia, {\em Hitting Probability for Anomalous Diffusion
Processes}, Phys. Rev. Lett., {\bf 104}, 020602 (2010).


\bibitem{RS2011} J. Rambeau, G. Schehr, {\em Distribution of the time at which $N$ vicious walkers
reach their maximal height}, Phys. Rev. E {\bf 83}, 061146 (2011).


\bibitem{DW2016} M. Delorme, K. J. Wiese, {\em Extreme-value statistics of fractional Brownian motion bridges}.
 Phys. Rev. E {\bf 94}, 052105 (2016).

\bibitem{AWW2020} M. Arutkin, B. Walter, K. J. Wiese,
{\em Extreme events for fractional Brownian motion with drift: Theory and numerical validation},
Phys. Rev. E, 102, 022102 (2020).

\bibitem{SP2021} P. Singh, A. Pal, {\em Extremal statistics for stochastic resetting systems},
Phys. Rev. E {\bf 103}, 052119 (2021).

\bibitem{Singh2020} P. Singh, {\em Extreme value statistics and arcsine laws for heterogeneous diffusion
processes}, Phys. Rev. E {\bf 105}, 024113 (2022).

\bibitem{MMS2021}F. Mori, S.~N. Majumdar, G. Schehr, {\em Distribution of the time of the maximum for
stationary processes}, Europhys. Lett. {\bf 135}, 30003 (2021).

\bibitem{MMS2022} F. Mori, S. N. Majumdar, G. Schehr, {\em Time to reach the maximum for a stationary stochastic 
process}, Phys. Rev. E {\bf 106}, 054110 (2022).

\bibitem{Zodage_2023}
A. Zodage, R. J. Allen, M. R. Evans, S. N. Majumdar,
{\em A sluggish random walk with subdiffusive spread},
J. Stat. Mech. 033211 (2023).


\bibitem{fa2003}
K. S. Fa, E. K. Lenzi,
\textit{Power law diffusion coefficient and anomalous diffusion: Analysis of solutions and first passage time},
Phys. Rev. E \textbf{67}, 061105 (2003).


\bibitem{cherstvy2013}
A. G. Cherstvy, A. V. Chechkin, R. Metzler,
\textit{Anomalous diffusion and ergodicity breaking in heterogeneous diffusion processes},
New J. Phys. \textbf{15}, 083039 (2013).

\bibitem{SCT2023} A. L. Stella, A. Chechkin, G. Teza, {\em Anomalous dynamical
scaling determines universal critical singularities}, Phys. Rev. Lett. {\bf 130},
207104 (2023).

\bibitem{SCT2023.1} A. L. Stella, A. Chechkin, G. Teza, {\em
Universal singularities of anomalous diffusion in the Richardson class},
Phys. Rev. E {\bf 107}, 054118 (2023).


\bibitem{watson_treatise_bessel}
G. N. Watson,
\textit{A Treatise on the Theory of Bessel Functions}, (Cambridge University Press, Cambridge, 1995).


\bibitem{sokhotski-plamelij-wiki}
{Wikipedia contributors},
\textit{Sokhotski–Plemelj theorem --- {Wikipedia}{,} The Free Encyclopedia},
2024, \url{https://en.wikipedia.org/w/index.php?title=
Sokhotski%E2%80%93Plemelj_theorem&oldid=1220081202} [Online; accessed 25-September-2024].


\bibitem{MC2005} S.N. Majumdar, A. Comtet, {\em Airy Distribution Function:
From the Area under a Brownian
Excursion to the Maximal Height of Fluctuating Interfaces},
J. Stat. Phys. {\bf 119}, 777 (2005).


\bibitem{PCMS13} A. Perret, A. Comtet, S. N. Majumdar, G. Schehr,
{\em Near-Extreme Statistics of Brownian Motion},
Phys. Rev. Lett., {\bf 111}, 240601 (2013).

\bibitem{PCMS15}
A. Perret, A. Comtet, S. N. Majumdar, G. Schehr, 
{\em On Certain Functionals of the Maximum of Brownian Motion and Their
Applications}, J. Stat. Phys., {\bf 161}, 1112 (2015).

\bibitem{CD2003} A. Comtet, J. Desbois {\em Brownian motion in wedges, last passage time and
the second arc-sine law}, J. Phys. A: Math. Gen. {\bf 36}, L255 (2003).



\end{thebibliography}
\end{document}